\documentclass[onecolumn]{aa}

\usepackage[varg]{txfonts}
\include{bookdefs}
\usepackage{graphicx}
\usepackage{xspace}
\usepackage{amsmath}
\usepackage{placeins}
\usepackage{gensymb}
\usepackage{color}
\usepackage{float}

\renewcommand{\vec}[1]{\mbox{\boldmath$#1$}}
\newcommand{\music}{\texttt{MUSIC}\xspace}

\definecolor{orange}{rgb}{.9,.3,0}

\begin{document}
\title{Extreme value statistics for two-dimensional convective penetration in a pre-Main Sequence star}
\author{J. Pratt\inst{\ref{inst1}}  \and I. Baraffe\inst{\ref{inst1},\ref{inst2},\ref{inst4}} \and  T. Goffrey\inst{\ref{inst1}} \and  T. Constantino\inst{\ref{inst1}}  \and M. Viallet\inst{\ref{inst3}} \and M.V. Popov\inst{\ref{inst2}}  \and R. Walder\inst{\ref{inst2}} \and D. Folini\inst{\ref{inst2}}   }
\institute{Astrophysics, College of Engineering, Mathematics and Physical Sciences, University of Exeter, EX4 4QL Exeter, United Kingdom \label{inst1} \and \'Ecole Normale Sup\'erieure de Lyon, CRAL (UMR CNRS 5574), Universit\'e de Lyon 1, 69007 Lyon, France\label{inst2} \and Max-Planck-Institut f\"ur Astrophysik, Karl Schwarzschild Strasse 1, 85741 Garching, Germany \label{inst3} \and Monash Centre for Astrophysics (MoCA) and School of Physics \& Astronomy, Monash University, Clayton Vic 3800, Australia \label{inst4}} 

\titlerunning{Extreme value statistics}
\authorrunning{J. Pratt et. al.}

\abstract
%\emph{Context:}
 {In the interior of stars, a convectively unstable zone typically borders a zone that is stable to convection.  Convective motions can penetrate the boundary between these zones, creating a layer characterized by intermittent convective mixing, and gradual erosion of the density and temperature stratification.}
%\emph{Aims:} 
{We examine a penetration layer formed between a central radiative zone and a large convection zone in the deep interior of a young low-mass star. Using the Multidimensional Stellar Implicit Code (\music) to simulate two-dimensional compressible stellar convection in a spherical geometry over long times, we produce statistics that characterize the extent and impact of convective penetration in this layer.}
%\emph{Methods:}   
{We apply extreme value theory to the maximal extent of convective penetration at any time.  We compare statistical results from simulations which treat non-local convection, throughout a large portion of the stellar radius, with simulations designed to treat local convection in a small region surrounding the penetration layer.   For each of these situations, we compare simulations of different resolution, which have different velocity magnitudes.  We also compare statistical results between simulations that radiate energy at a constant rate to those that allow energy to radiate from the stellar surface according to the local surface temperature.}
%\emph{Results:} 
{Based on the frequency and depth of penetrating convective structures, we observe two distinct layers that form between the convection zone and the stable radiative zone.  
We show that the probability density function of the maximal depth of convective penetration at any time corresponds closely in space with the radial position where internal waves are excited.   We find that the maximal penetration depth can be modeled by a Weibull distribution with a small shape parameter.   Using these results, and building on established scalings
for diffusion enhanced by large-scale convective motions, we propose a new form for the diffusion coefficient that may be used for one-dimensional stellar evolution calculations in the large P\'eclet number regime.  These results should contribute to the 321D link.}
{}

\maketitle

\section{Introduction}

Convection is a fundamental stellar process which underlies heat transport, mixing, shear, and the dynamo.  
During many phases of stellar evolution, a star's interior is characterized by layers that are convectively
unstable.  Outside of a convection zone are layers that are stable to convection.  Convective fluid motions can cross the boundary between a convectively unstable region and a neighboring stable region.  When the P\'eclet number is large, these convective motions slowly erode the density and thermal stratification of the stable region.  When this occurs, the process is termed convective penetration.  We study convective motions that may contribute to convective penetration below a convection zone, and thus can have an impact on the structure and evolution of a star. 

Studies of stellar evolution utilize one-dimensional calculations that evolve physical quantities as a function of the radial position interior to a star.  
The effects of convection on other physical quantities are typically modeled using stellar mixing length theory, which depends on the local temperature gradient \citep[e.g.][]{vitense1953wasserstoffkonvektionszone, bohm1958wasserstoffkonvektionszone,abbett1997solar,trampedach2010convection,brandenburg2015stellar}.  
The phenomenon of convective penetration has been included in mixing length theory in two competing ways, based on different non-local convection theories.  One is an overshooting length, expressed as a fraction of the pressure scale height at the convective boundary, which defines a region outside of the convection zone where the stellar model assumes full convective mixing.  The other is a region of enhanced one-dimensional diffusivity \citep{noels2010overshooting,zhang2013convective}, where the intensity of convective mixing changes with radius.  
Using statistical observations, this work addresses both of these methods and thereby contributes to the 321D link.

Stellar convection is a nonlinear and non-local phenomenon.  The extent to which convective motions can extend beyond the theoretical boundary of a convection zone is intuitively linked to the local velocity, density, and temperature of the fluid at the boundary\footnote{The local interplay of velocity, density, and temperature have been related to buoyancy breaking of plumes \citep[e.g.][]{latour1981stellar,hughes1988magnetic,spruit1990solar}.}.  In this work we will refer to convective flow structures that possess radial velocity as plumes, without invoking geometry or physics specific to plume studies.  Plumes frequently move a short distance beyond the convective boundary.  Less frequently, plumes penetrate the stable zone by a larger extent.    These plumes that extend deeper clearly have a greater effect on the star, because they penetrate with larger momentum and affect the stratification deeper inside the stable zone.   
Convective penetration is therefore naturally a process in which the intermittency is significant. 
A theoretical description of such flows should be built upon statistics gathered by observing convective penetration over long times.  The Multidimensional Stellar Implicit Code (\music), which uses implicit time integration, allows us to cover efficiently the long times necessary for converged statistics on convective penetration.  This work thus builds on earlier explorations \citep[e.g.][]{hurlburt1986nonlinear,hurlburt1994penetration,brummell2002penetration,rogers2005penetrative,rogers2006numerical}, which simulate convective overshooting or penetration over shorter times, typically 2-5 convective turnover times.

In addition to observing the long-time behavior of convective penetration, a statistical treatment is necessary to model meaningfully instances of large convective penetration.  Extreme value theory (EVT) is a statistical theory that has been widely used to predict events that have large impacts. Examples are record rainfall, floods, dangerously high winds, earthquakes, and avalanches.   For the case of convective flows penetrating a stable zone, an analysis using extreme value theory is natural.
  
This work is structured as follows.  In Section~\ref{secsim} we discuss the young sun model and the numerical framework of our simulations.  In Section~\ref{secextent} we examine convective penetration based on simulation data that covers a range of times, and a range of different spatial resolutions and extents.
Motivated by these results, in Section~\ref{secevd} we introduce the framework of extreme value theory, and compare statistical results 
for simulations in a truncated domain that model penetration due to local convection, and simulations in a large domain that model penetration due to non-local convection.
  In Section~\ref{secdiffcoef} we build on these statistical results to formulate
a one-dimensional radial diffusion coefficient that is relevant in the large P\'eclet number limit.
In Section~\ref{secfinal} we summarize and discuss the implications of these results.

\section{Simulations \label{secsim}}

In this work we examine a model of a prototypical pre-main sequence star, the young sun, using the numerical set-up benchmarked in \citet{goffrey2016benchmarking}.   In \citet{pratt2016spherical} the structure and properties of the stellar model of the young sun are described in detail.  Here we briefly summarize the salient features of our simulations.
We perform two-dimensional implicit large eddy simulations (LES) of the young sun using the \music code.  The stellar radius of our young sun model is approximately three times larger than the current sun, it is one solar mass, and has homogeneous chemical composition.  The radial profiles of density and temperature for the young sun model are typical for a pre-main sequence star that is no longer accreting and is gradually contracting.  The luminosity is increasing with the interior radius.   Based on the radial entropy profile and an evaluation of the Schwarzschild criterion, a central radiative zone is expected below a large convection zone.  This convection zone spans the outer $1.2 \cdot 10^{11}$ cm of the total radius of $2.13 \cdot 10^{11}$ cm.    A layer exists between the convection zone and the radiative zone, where mixing of convective flows into the radiative zone can take place.
This layer is the focus of our work, and is located far from the physically complicated near-surface layers.   
In this layer, we estimate the stiffness \citep[e.g.][]{hurlburt1994penetration,brummell2002penetration,rogers2006numerical} to be $S \approx 13$.
Our simulations do not include additional stellar physics such as rotation, a tachocline, chemical mixing, or magnetic fields, that we intentionally omit from our statistical study of convective penetration. This simplifies the dynamics at the boundary between the interior radiative zone and the convection zone, where we observe penetrative convective motions.   These additional effects may influence the results we obtain.  In three dimensions the shape of plumes may be different from two dimensions \citep{schmalzl2004validity,van2015comparison}, and this could affect mixing properties.  Rotation and shear may tilt plumes, modifying the angle of penetration \citep[e.g.][]{brun2017differential}.  A magnetic field may shift the Schwarzschild criteria \citep{gough1966influence} and fundamentally change the mechanisms of plume braking \citep{hughes1988magnetic}. These additional physical effects will be explored in future work. 

The \music code solves the inviscid compressible hydrodynamic equations for density $\rho$, momentum $\rho \vec{u}$, and internal energy $\rho e$:
\begin{eqnarray} \label{densityeq}
\frac{\partial}{\partial t} \rho &=& -\nabla \cdot (\rho \vec{u})~,
\\ \label{momeq}
\frac{\partial}{\partial t} \rho \vec{u} &=& -\nabla \cdot (\rho \vec{u} \vec{u}) - \nabla p + \rho \vec{g} ~,
\\ \label{ieneq}
\frac{\partial}{\partial t} \rho e &=& -\nabla \cdot (\rho e\vec{u}) + p \nabla \cdot \vec{u} + \nabla \cdot (\chi \nabla T) .
\end{eqnarray}
using a finite-volume method, a MUSCL scheme for interpolation \citep{van1977towards}, and a van Leer flux limiter \citep{van1974towards,roe1986characteristic}.
Time integration in the \music code is fully implicit, and uses a Jacobian free Newton-Krylov (JFNK) solver \citep{knoll2004jacobian} with a variant of physics-based preconditioning \citep{maximepaper}.  For all simulations examined in this work, an adaptive time-step maintains a general CFL number of 10, while a CFL number based on simple advection is restricted to be 0.5. We find that this produces good convergence of relevant basic quantities, such as the average kinetic energy. 

\music simulations are designed to contribute to the 321D link \citep{david20143d, arnett2009turbulent}, \emph{i.e.} the effort to improve one-dimensional stellar evolution models by studying critical short phases using stellar hydrodynamics in two and three dimensions.
One aspect that makes our simulations relevant to the 321D link is the use of an equation of state and realistic opacities standard in one-dimensional stellar evolution calculations. Opacities are interpolated from the OPAL \citep{iglesias1996updated} and \citet{ferguson2005low} tables, which cover a range in temperature suitable for the description of the entire structure of a low-mass star. The compressible hydrodynamic equations \eqref{densityeq}-\eqref{ieneq} are closed by determining the pressure $p(\rho,e)$ and temperature $T(\rho,e)$ from a tabulated equation of state for a solar composition mixture.
This equation of state accounts for partial ionization of atomic species by solving the Saha equation, and neglects partial degeneracy of electrons; it is suitable for the description of our solar model at a young age. The initial state for \music simulations is produced using data extracted from a one-dimensional model calculated using the Lyon stellar evolution code \citep{baraffe1991evolution, baraffe1997evolutionary,baraffe1998evolutionary}, which uses the same opacities and equation of state implemented in \music.
 In eq. \eqref{momeq}, $\vec{g}$ is the gravitational acceleration, a spherically-symmetric vector identical to that used in the stellar evolution code, and not evolved in our simulations.  

The thermal diffusivity in \music is also realistic for the young sun model, and has not been enhanced.   In eq. \eqref{ieneq} the thermal conductivity $\chi=16 \sigma T^3/3 \bar{\kappa} \rho$ is defined using  the Stefan-Boltzmann constant $\sigma$ and the Rosseland mean opacity  $\bar{\kappa}$.
An artificially high thermal diffusivity is a key issue that has been found to cause differences between mixing length theory and numerical results
\citep{rempel2004overshoot}, and can result in structural changes for a star.  Despite these consequences, enhancing the thermal diffusivity has become a conventional step in numerical studies of convective overshooting \citep[][]{browning2004simulations,rogers2006numerical,tian2009numerical,strugarek2011magnetic}, because it allows the star to more rapidly achieve energy balance by shortening
the Kelvin-Helmholtz time scale.  For the young sun model considered here, the Kelvin-Helmholtz time scale is several orders of magnitude greater than its convective turnover
time.  Thus although the young sun model has not achieved energy balance, during the time that our simulations span, there
is no measured change in the thermal profile, and the convection may be considered to be in steady state.

When a realistically small thermal diffusivity is used, a high resolution is typically required to resolve the corresponding small-scale temperature
fluctuations.    We do not attempt to fully resolve the small scales related to the realistically small thermal diffusivity that we use; we rely on the implicit large eddy simulation paradigm to approximate the effect of this small-scale dissipation.  Our use of the realistic thermal diffusivity for the young sun model allows us to study convective penetration because it produces a large P\'eclet number; this is discussed fully in Section \ref{secpeclet}. In \music, numerical truncation errors contribute to both the viscosity and thermal diffusivity.  Because \music simulations also use an explicit thermal diffusivity related to the thermal conductivity in eq. \eqref{ieneq}, the Prandtl number is expected to be everywhere less than one.
The characteristic length scales, velocities, viscosity, and thermal diffusivity vary throughout the radius of a star.  The viscosity and thermal diffusivity may also be dependent on local dynamics, the wavenumber, and the estimation method \citep{domaradzki2003effective,zhou2014estimating,radice2015implicit,schranner2015assessing, zhou2016comparison}.  For these reasons, the Rayleigh number and Reynolds number are not specified in a general sense for our simulations.

\subsection{Spherical-shell geometry and boundary conditions \label{secbc}}
    
The compressible hydrodynamic equations \eqref{densityeq}-\eqref{ieneq} are solved in a two-dimensional spherical shell using spherical coordinates:  radius $r$ and colatitude $\theta$.  A two-dimensional geometry has been chosen in order to allow us to produce a long time series of data that also has a satisfactory resolution near the bottom of the convection zone.   
Convective penetration is the result of large-scale convection, and our simulations do not model the significantly three-dimensional effects of turbulence, rotation, or magnetic fields present in a realistic star.  Therefore a two-dimensional simulation provides an approximation to three-dimensional stellar convection.  It has long been recognized that two-dimensional stellar simulations result in higher velocity magnitudes than three-dimensional simulations \citep{muthsam1995numerical,meakin2007turbulent}.  However, because our simulations do not approach the high velocities realistic to a star, the difference between two- and three-dimensional velocities do not seriously detract from a study of the form of convective penetration.

The placement of boundaries and the choice of boundary conditions affect the physical outcome of
a hydrodynamic simulation.  We use boundary conditions targeted to maintain the original radial profiles of density and temperature of the one-dimensional stellar evolution model of the young sun. 
Each simulation volume begins at $20 \degree$ from the north pole, and ends $20 \degree$ before the south pole, allowing for a wide convection zone with multiple descending and ascending plumes.  We impose periodicity on all physical quantities at the boundaries in $\theta$.  Radial boundaries require more sophisticated treatment.  In velocity, we impose non-penetrative and stress-free boundary conditions on the radial boundaries.  The energy flux and luminosity are constant at the inner radial boundary at the value of the energy flux at that radius in the one-dimensional stellar evolution calculation.  Typically we also hold the energy flux and surface luminosity constant on the outer radial boundary as well; the surface luminosity of our model for the young sun is 2.32 times the luminosity of the current sun.    An alternative is to allow the surface to radiate energy with the local surface temperature.  In this case the energy flux  varies as $\sigma T_{\mathsf{s}}^4$ where $\sigma$ is the Stefan-Boltzmann constant and $T_{\mathsf{s}}(\theta,t)$ is the temperature along the surface.  This boundary condition can only be effectively used when the temperature gradient near the surface is sufficiently resolved; otherwise it results in artificially high cooling rates. 
On the inner radial boundary of the spherical shell, we impose a constant radial derivative on the density, as discussed in \citet{pratt2016spherical}.  
  At the outer radial boundary we apply a hydrostatic equilibrium boundary condition on the density that maintains hydrostatic equilibrium by assuming constant internal energy and constant radial acceleration due to gravity in the boundary cells \citep{hsegrimm2015realistic}.

We consider the suite of two-dimensional simulations of spherical-shell convection summarized in Table~\ref{simsuma}.    
  For all of these simulations, the table lists radial grid spacing and the radial extent of the spherical shell in units of the total stellar radius $R$. 
\citet{rogers2006numerical} report using a grid of $N_r \times N_{\theta} = 2048 \times 1500$, with 620 grid cells in the radiative zone to study convective overshooting in the current sun.  By comparison, in simulation YS5 we use a grid of $N_r \times N_{\theta} = 2432 \times 2048$ with 922 grid cells in the radiative zone to study convective penetration in the young sun.  The large section of the radiative zone simulated at high resolution is useful to precisely resolve convective penetration in the young sun, which has a larger convection zone than the current sun.

\begin{figure*}[h]
\begin{center}
\resizebox{2in}{!}{\includegraphics{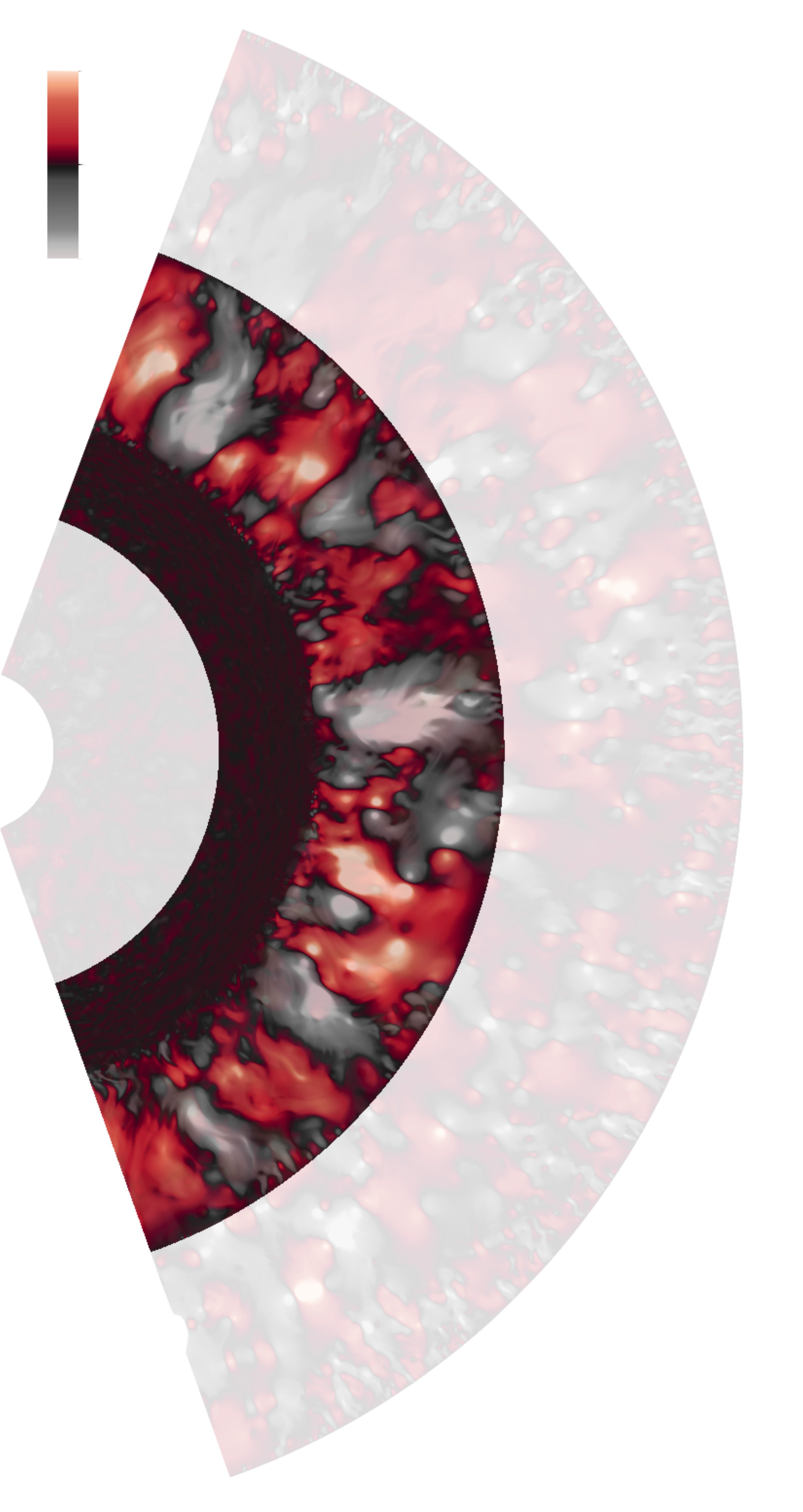}}\resizebox{2.2in}{!}{\includegraphics{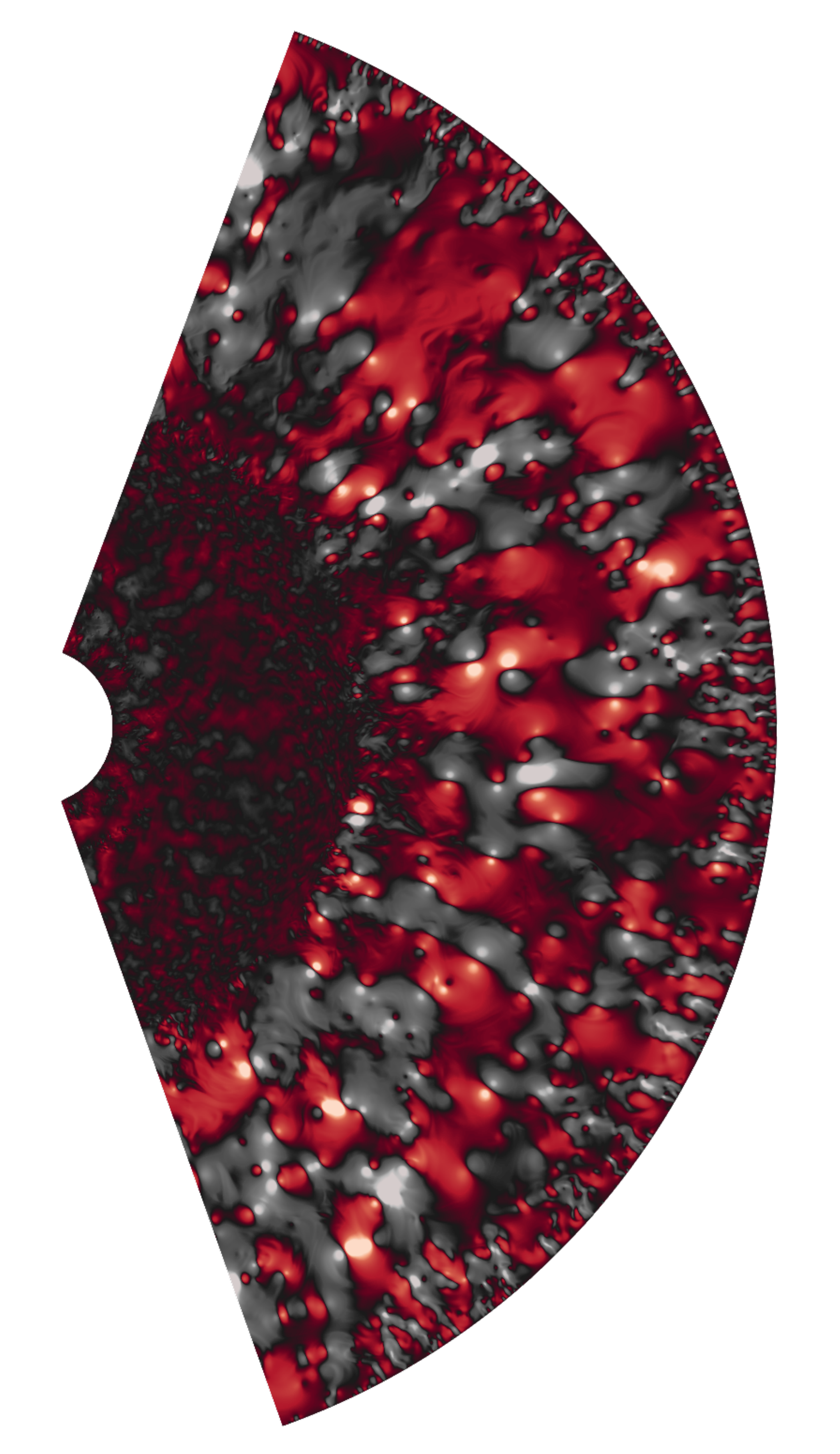}}\resizebox{2.15in}{!}{\includegraphics{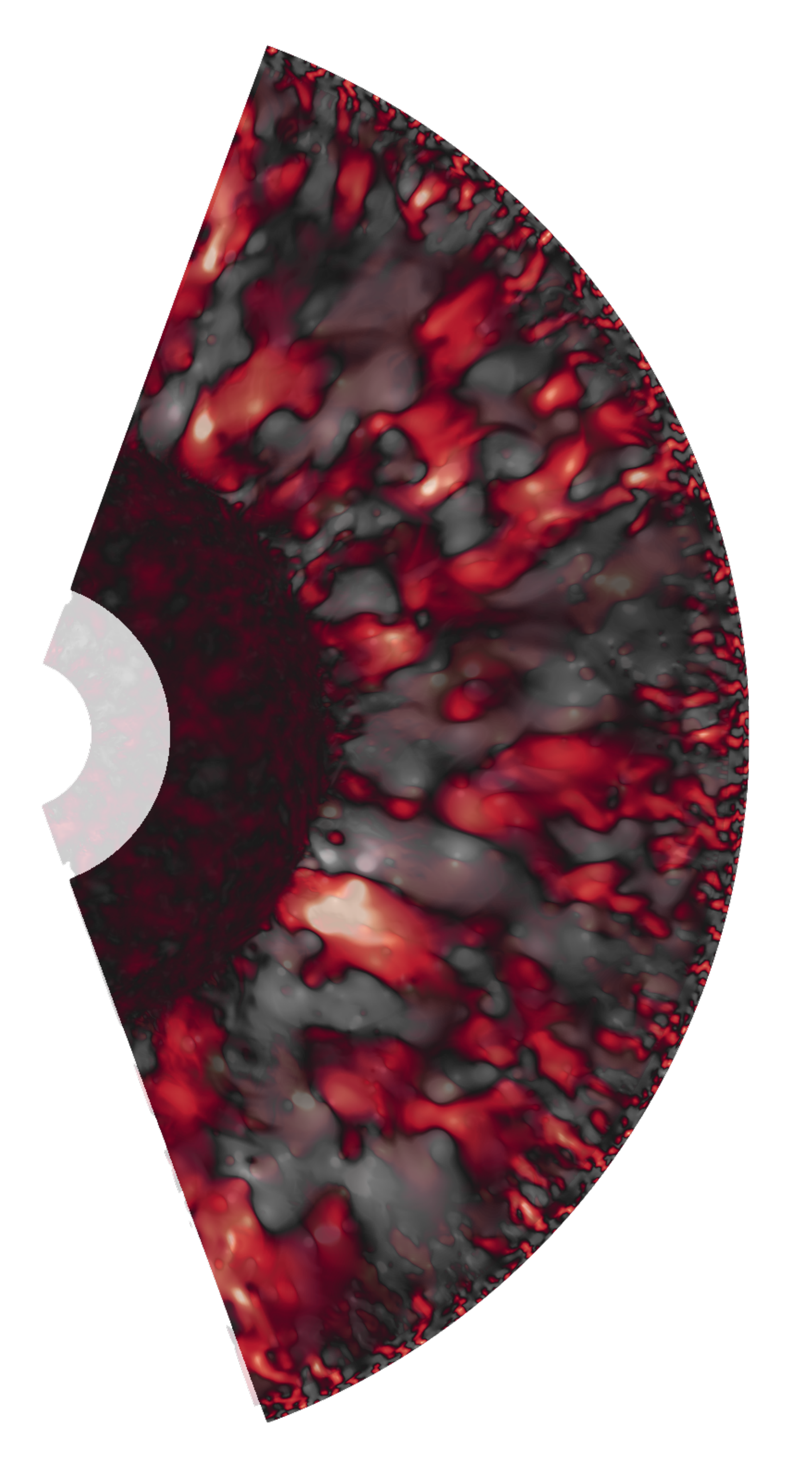}}
\caption{Typical time snapshots of radial velocity in simulation YS2 (left), YS5 (middle), and YS6 (right).  The zero point is colored black, while inward flows are grey and outward flows are red.
\label{vizbox}}
\end{center}
\end{figure*}

Table~\ref{simsuma} provides the parameters of simulations YS0-6, which consist of three different sets of simulations.  Simulations YS0-2 model convection in the local approximation, using a spherical shell of limited radial extent around the penetration layer.  These simulations are designed to examine penetration due to
convection that arises and is dynamically limited to the layer around the bottom of the convection zone, similar to a box-in-a-star approach.
  Simulations YS3-5 model convection in the non-local approximation, using a spherical shell that includes nearly the whole star, terminating in the near-surface layers.
 These simulations are designed to examine penetration due to
convection that is dynamically accurate to the large convection zone of the young sun model, where convective dynamics may interact non-locally across more than half of the stellar radius. 
    Among each of these sets of simulations, three different resolutions are examined for otherwise identical physical set-ups and for boundary conditions that include a constant energy flux at the outer surface.  In \music simulations, which do not include an explicit viscosity, higher resolution produces higher characteristic velocities in the convection zone.  Thus these different resolution simulations are performed to show the robustness of our statistical results with respect to convective velocities.  An additional simulation YS6 is performed in order to isolate the effect of allowing the energy flux at the stellar surface to vary according to the local temperature.  A typical snapshot of the radial velocity in each of these simulation volumes is shown in Fig.~\ref{vizbox}.  The effect of the placement of the simulation boundaries and the boundary conditions imposed on the properties of convection was examined in \citet{pratt2016spherical}.  In this work we examine exclusively the penetration of plumes below the convection zone.

For each simulation, Table~\ref{simsuma} provides a time span in units of the convective turnover time of that simulation.  This time span indicates a period after steady convection has developed, during which the global kinetic energy of the system is approximately constant. The average radial structure of the star's density and temperature does not evolve over this period.  All time-averages and other statistics are calculated over this time span.

 \begin{table*}
\begin{center}
\caption{Parameters for two-dimensional compressible hydrodynamic simulations of the young sun.
 \label{simsuma}
 }
\begin{tabular}{lccccccccccccccccccccccccccc}
\hline\hline
                      & $R_{\mathsf{in}}$  & $R_{\mathsf{out}}$  & $\Delta r$ ($10^{-4} /R$) & $L_{\mathsf{in}}$  & $L_{\mathsf{out}}$&  $\tau_{\mathsf{conv}}$($10^6$s) & time span ($\tau_{\mathsf{conv}}$) & surface radiation %

\\ \hline
Local:
\\ \hline
YS0                 & 0.31        & 0.67   & 14 & $5.82 \cdot 10^{-2}$ & 1.93 &  $ 2.52 \pm 0.60 $& 525  & constant % 
\\ \hline
YS1                  & 0.31        & 0.67   & 7 &  $5.82 \cdot 10^{-2}$ &1.93 &  $ 3.31 \pm 0.36 $& 152  & constant % 
\\ \hline
YS2                  & 0.31         & 0.67   &  3.5 & $5.82 \cdot 10^{-2}$ & 1.93 & $ 3.59 \pm 0.37 $  & 35  & constant  %
\\ \hline\hline
Non-local:
\\ \hline
YS3                 & 0.10          & 0.97     & 14  & $5.37 \cdot 10^{-3}$ & 2.32 &  $ 1.19 \pm 0.15 $  & 143  & constant %
\\ \hline 
YS4                  & 0.10         & 0.97    & 7  & $5.37 \cdot 10^{-3}$ & 2.32&  $ 1.22 \pm 0.21 $  &  71  & constant % 
\\ \hline 
YS5                  & 0.10         & 0.97    & 3.5  & $5.37 \cdot 10^{-3}$ & 2.32&  $ 1.03 \pm 0.05 $  &  5  & constant % 
\\ \hline\hline
YS6                 & 0.21         & 1.00     &14 &  $3.36 \cdot 10^{-2}$ & 2.32 &  $ 0.67 \pm 0.09 $  & 197  & varying %
\\ \hline\hline
\end{tabular}
\tablefoot{The inner and outer radius of the spherical shell, and the radial resolution in the convection zone are given in units of the total radius of the young sun $R$.
The luminosity on the inner and outer radii are given in units of the solar luminosity, $3.839 \cdot 10^{33} \mathsf{erg/s}$.
  The convective turnover time $\tau_{\mathsf{conv}}$, and the total time span for each simulation is summarized.  The method of radiation of internal energy at the surface is also indicated.   Simulations that test set-ups for local convection and non-local convection are included.
}
\end{center}
\end{table*}

 \subsection{Characterization based on the P\'eclet number \label{secpeclet}}

%bookmark
The P\'eclet number is a nondimensional ratio that measures the relative importance of two physical effects that determine the deceleration of a fluid element
entering a layer that is stable to convection: advection and thermal diffusion.  Therefore it is standardly used to distinguish between convective overshooting and convective penetration.  Low P\'eclet numbers ($\mathsf{Pe} \leq 1$) indicate that thermal
 diffusion is the dominant process and convective motions overshoot the convection zone without significantly changing the stratification of the stable layer; high P\'eclet numbers indicate penetration can take place, a process where convective fluid motions modify the thermal and density stratification in the stable layer outside the convection zone \citep[e.g.][]{zahn1991convective}. The P\'eclet number is typically defined
\begin{eqnarray}\label{eqpecdef}
\mathsf{Pe}=v_{\mathsf{RMS}} h_p/\alpha ~,
\\
\alpha= \chi/\rho c_{\mathsf{P}}~.
\end{eqnarray}
Here the pressure scale height $h_p$ represents a characteristic length scale, and the root-mean-square velocity $v_{\mathsf{RMS}}$ represents a characteristic convective velocity.  The thermal diffusivity $\alpha$ is produced from our tabulated equation of state, using the definition of the thermal conductivity
$\chi$ in eq.~\eqref{ieneq}, and the specific heat capacity at constant pressure $c_{\mathsf{P}}$.    Numerical truncation errors due to the finite resolution of a simulation can increase this thermal diffusivity.

\begin{figure}[h]
\begin{center}
\resizebox{3.5in}{!}{\includegraphics{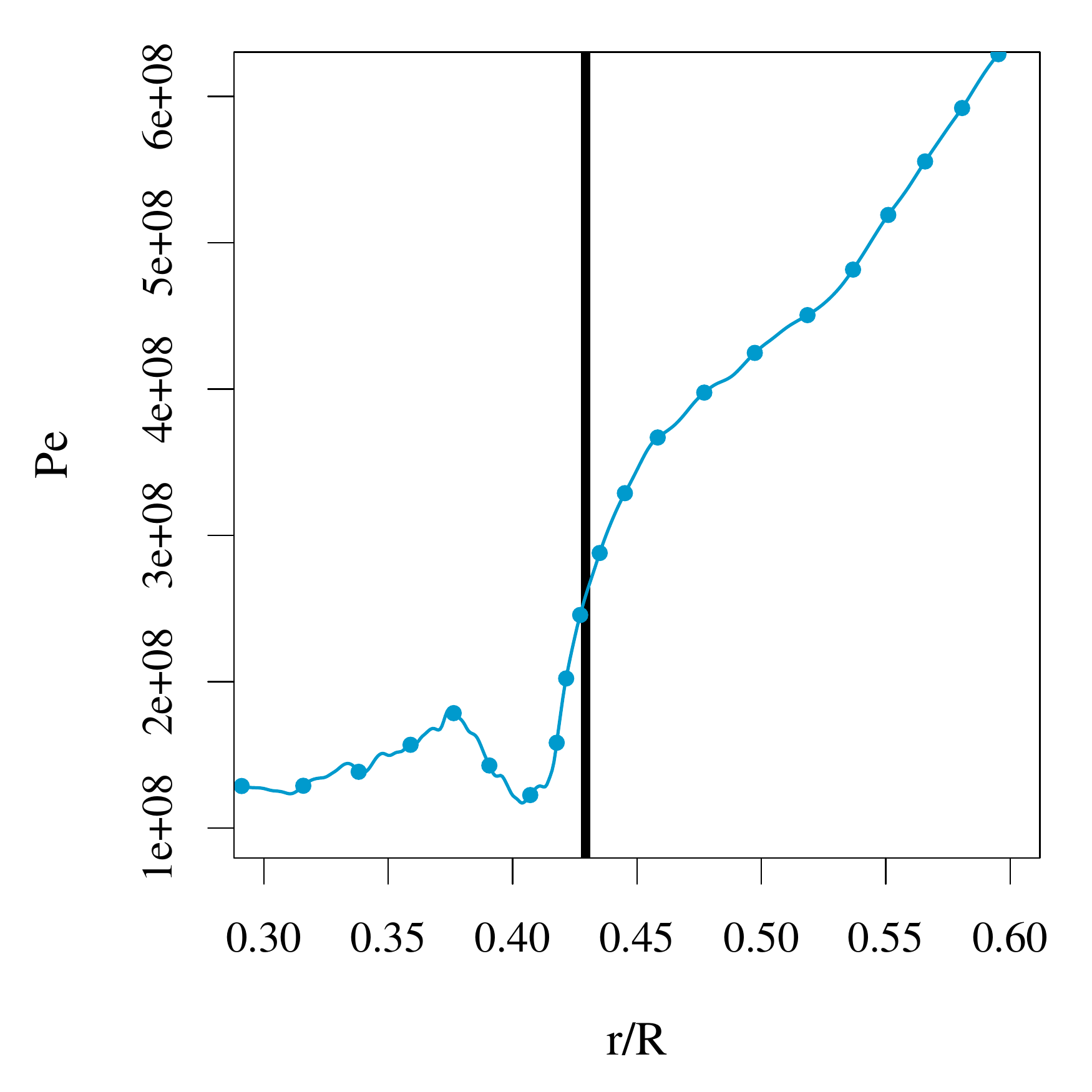}}
\caption{Radial profile of the time-averaged and horizontally-volume-averaged P\'eclet number, as defined by eq.~\eqref{eqpecdef}, from simulation YS4.  The contribution to the thermal diffusivity by numerical truncations are not included in this profile.  The heavy vertical black line marks the boundary between the stable radiative zone and the convection zone, determined from the radial profile of entropy and the Schwarzschild criterion produced by the one-dimensional stellar model.
\label{pecplot}}
\end{center}
\end{figure}

In our simulations, radial profiles of temperature and thermal diffusivity are taken directly from the model of the young sun produced by a one-dimensional stellar evolution calculation.  Because of this, and because we use a realistically small thermal diffusivity, the P\'eclet number is much greater than 1 in the area of interest at the bottom of the convection zone, and convective penetration is expected.  Fig.~\ref{pecplot} shows the average radial profile of the P\'eclet number in simulation YS4.  
Neglecting the contributions to thermal diffusivity from numerical truncation, the P\'eclet number is $\mathcal{O}(10^7)$ -- $\mathcal{O}(10^8)$ in the layer at the bottom of the convection zone for all of the simulations in Table~\ref{simsuma}.
By perturbing the background state we are able to measure an approximate effective thermal diffusivity. We find that while numerical truncation errors increase the effective thermal diffusivity beyond the value implied by the explicit physical thermal diffusivity, our simulations are still in the $\mathsf{Pe} \gg 1$ regime.
 A P\'eclet number larger than one is expected; P\'eclet numbers are generally higher in low and medium mass stars, like the young sun model,
than for higher mass stars \citep[e.g.][]{meynet2000stellar}.  P\'eclet numbers are also expected to be much higher in stellar interiors than in the near-surface layers.  In addition to these physical points, two-dimensional hydrodynamic simulations are known for producing somewhat higher P\'eclet numbers because they exhibit higher velocities than three dimensional simulations \citep{muthsam1995numerical,meakin2007turbulent}.

Based on the P\'eclet number, we define convective motions that cross beyond the boundary of the convection zone to be convective penetration.  Indeed in test simulations where the Kelvin-Helmholtz time scale of the young sun is artificially lowered by the use of an enhanced thermal diffusivity and corresponding increased luminosity, we observe penetrative motions that change the density and thermal stratification of the star.  However in the simulations presented in this work, the Kelvin-Helmholtz time scale is many orders of magnitude larger than the characteristic time scale associated with convective motions\footnote{The time-scale for thermal evolution of the young sun model estimated to be on the order of 4 Myr, or $\mathcal{O}(10^8) ~\tau_{\mathsf{conv}}$.}.  We therefore do not observe erosion of the stellar structure due to convective penetration; the temperature and density profiles are approximately constant during our simulations.  We do study penetrative flows that have the potential to change stellar structure.  However our study may also have implications for convective overshooting motions, i.e. convection in the low P\'eclet number regime.

Over the last decade it has become common to refer to the mixing at the boundary of a convection zone as turbulent entrainment \citep[e.g.][]{meakin2007turbulent}.  We note 
that convective penetration has long been considered a type of entrainment in which mixing is due to large-scale convective motions \citep[e.g.][]{tennekes1981basic}, 
and that in a physical star, mixing at a boundary may be due to a combination of fluid effects including convection, turbulence, and also shear \citep[e.g.][]{jonker2013scaling}.  Generally in LES of stellar convection, the range of scales that correspond to turbulent mixing include scales considerably smaller than the grid spacing of the simulation.  Indeed for all of the simulations considered in this work, flows are expected to be in the laminar regime with moderate Reynolds numbers, and the effect of turbulent entrainment is not targeted.

\section{Results: The penetration depth \label{secextent}}

\subsection{Determination of the extent of convective penetration \label{secdefovershoot}}

Historically several different definitions have been used for the depth of convective penetration.   Broadly speaking, a physical quantity is sought that either drops to a low level, or changes sign where convective motions cease.  To measure the penetration depth using a quantity that drops to a low level,  an arbitrary percentage of the quantity is chosen.  An example of such a quantity is the kinetic energy density $\frac{1}{2}\rho \vec{v^2}$.  In several earlier works, the point where convective motions cease has been defined as the point where the kinetic energy density reaches 1\%  \citep{brummell2002penetration}, or 5\%  \citep{rogers2006numerical} of its peak value near the bottom of the convection zone.

\begin{figure}[h]
\begin{center}
\resizebox{3.5in}{!}{\includegraphics{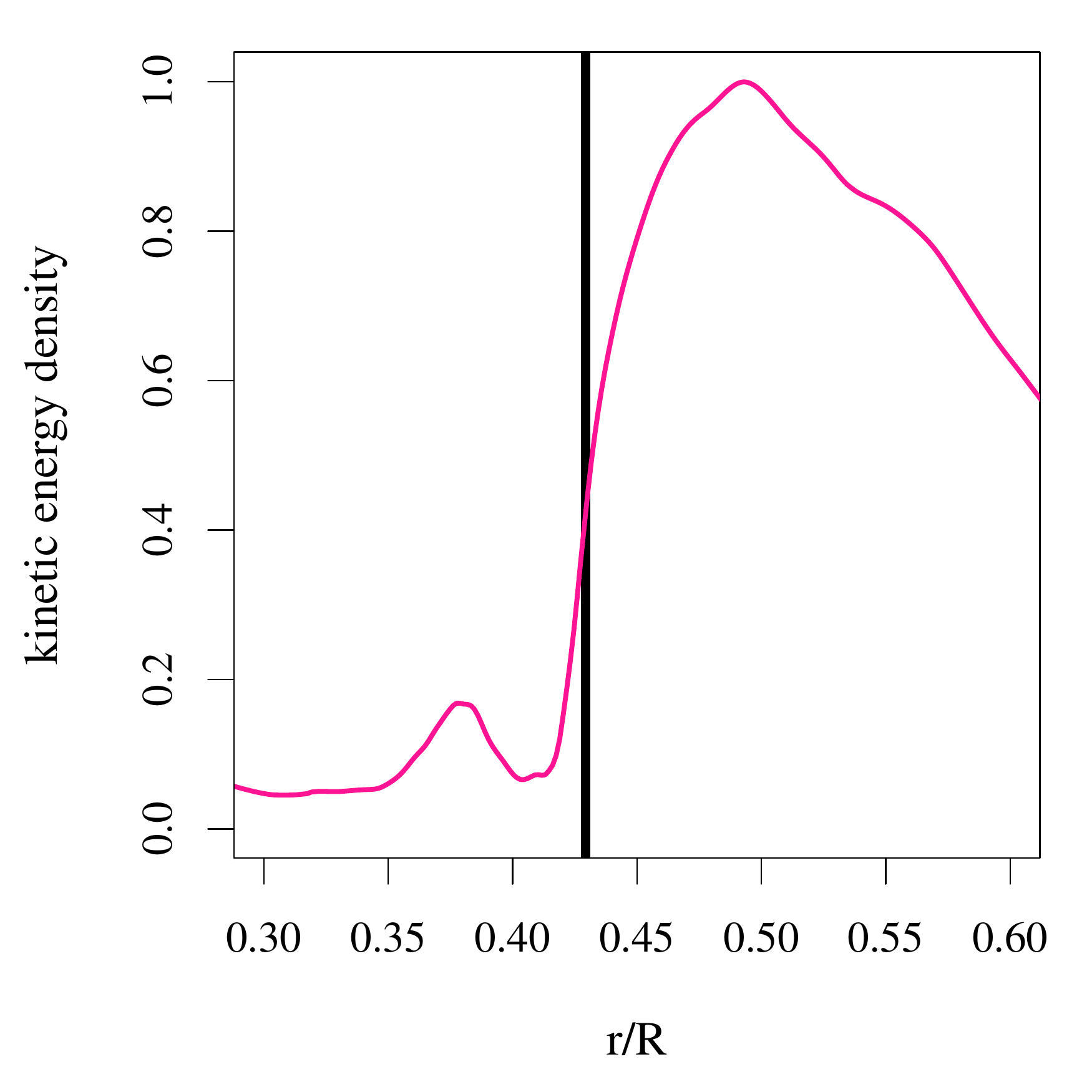}}
\caption{Radial profile of the time-averaged and horizontally volume-averaged kinetic energy, $\frac{1}{2}\rho \vec{v}^2$, scaled to its maximum, from simulation YS3.  The heavy vertical black line marks the convective boundary.
\label{keplot}}
\end{center}
\end{figure}

In our simulations, difficulty with this type of measurement arises after convective plumes interact with the underlying radiative region.  Convective penetration and overshooting generate a spectrum of waves in the radiative zone that propagate in the angular direction and feed back on the convective plumes.  These waves have been identified as internal gravity waves in many cases \citep{hurlburt1986nonlinear,andersen1994excitation,dintrans2003stochastic,rogers2005gravity,rogers2006numerical,lecoanet2013internal,shiode2013observational,brun2013gravity,alvan2014theoretical,alvan2015characterizing,pinccon2016generation}.   We observe waves in the radiative zone evolving over tens of convective turnover times, and they spoil an evaluation of a percentage of the peak convection zone kinetic energy density.  This problem is clear from the long-time-average of the kinetic energy density shown in Fig.~\ref{keplot} for simulation YS3.   In this long-time average, these waves appear as a small peak in the kinetic energy density at a radius $0.35<r/R<0.4$, precisely in the region where plumes that have a large penetration depth terminate.    A second issue arises when we observe steady convection over tens of convective turnover times: 
 the peak value of the kinetic energy density near the bottom of the convection zone evolves.  Thus a criterion based on a percentage of the peak value changes during the long time span of our simulations.     

\begin{figure}[h]
\begin{center}
\resizebox{3.5in}{!}{\includegraphics{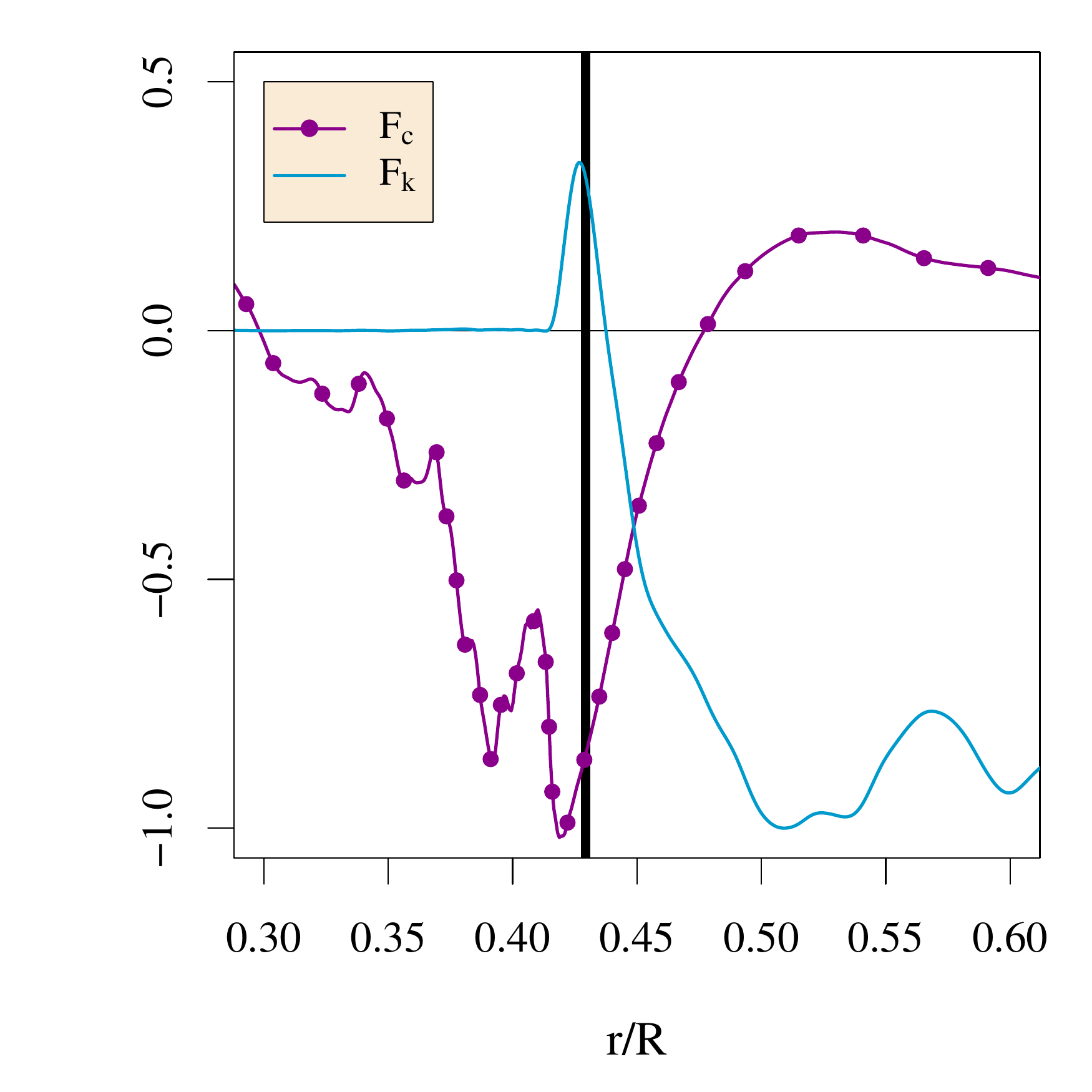}}
\caption{ Radial profile of the time-averaged and horizontally volume-averaged  
vertical kinetic energy flux $F_k$ defined in eq.~\eqref{kfluxeq}, and vertical heat flux $F_c$ defined in eq.~\eqref{cfluxeq} for simulation YS4.  Both of these fluxes have been normalized to their absolute maximum value.    The heavy vertical black line marks the convective boundary.
\label{kfluxplot}}
\end{center}
\end{figure}

For these reasons it is advantageous to base the penetration depth on a quantity that changes sign at the point in the radiative zone where convective motions cease.  Two such quantities have been proposed \citep[as discussed in, e.g.][]{hurlburt1986nonlinear, hurlburt1994penetration, ziegler2003box, rogers2006numerical,tian2009numerical, chan2010overshooting}: the vertical kinetic energy flux and the vertical heat flux.  We examine both of these measures, and compare the results.  The vertical kinetic energy flux $F_k$ is defined as
\begin{eqnarray}\label{kfluxeq}
F_k=\frac{1}{2}\vec{v}_r \rho \vec{v}^2 ~,
\end{eqnarray}
where $\rho$ is the density, and $\vec{v}$ is the velocity.
The penetration layer can be defined by the region of positive vertical kinetic energy flux. 
A radial profile of the vertical kinetic energy flux, calculated by averaging both in time and using a volume weighted-average in the horizontal ($\theta$) direction, is shown in Fig. \ref{kfluxplot} for simulation YS4. Previous works have used this type of averaged profile to determine the penetration depth.  In Fig. \ref{kfluxplot}, the vertical kinetic energy flux exhibits a 
smooth positive peak; the width of the layer where convective mixing takes place is well defined, but small.  In any time-snapshot, larger penetration may be identified at particular angles.

The second criteria used to measure the penetration depth is the vertical heat flux:
\begin{eqnarray}\label{cfluxeq}
F_c = \rho \vec{v}_r(\delta T)c_P~,
\end{eqnarray}
where the temperature perturbation $\delta T$ is defined relative to a time and horizontally volume-averaged temperature profile.  
Although the time and horizontally volume-averaged temperature profile is allowed to evolve during our simulations, it does not change substantially; this simplifies the calculation of the temperature perturbation.
A layer defined by convective penetration is associated with negative vertical heat flux. 
Fig.~\ref{kfluxplot} also shows the time-averaged and horizontally volume-averaged vertical heat flux for simulation YS4.   The penetration layer is clearly identifiable as the negative peak of vertical heat flux surrounding the bottom of the convection zone.  
There is significant discrepancy between the penetration depth defined from the \emph{averaged} vertical kinetic energy flux and the \emph{averaged}  vertical heat flux.
 Because of this, we examine both measures in detail.  We note that the use of Lagrangian coherent structures \citep[e.g.][]{hadjighasem2015spectral} to define the shape of convective plumes offers an alternative way to understand this discrepancy.

Horizontal and time averages have been used historically to define the width of the penetration layer. However the full statistics of convective penetration have not been previously examined, nor has the use of a simple average to represent them been specifically justified.  Straightforward averages can mask contributions from the intermittent convective plumes that penetrate deeper into the stable radiative zone than the average plume.  The distribution of penetration lengths may not be symmetric, or may have other relevant features that are not captured by a mean.  Just such intermittent plumes can have a potentially larger impact on the physics of the penetration layer.  Indeed \citet{christensen2011more} find that consideration of temporal and spatial inhomogeneity of overshooting could lead to a thermal profile that agrees with observations from helioseismology.  We therefore approach the full statistics of penetrating plumes with the goal of quantifying the contributions of intermittency.

\subsection{Shape of the penetration layer}

\begin{figure*}
\begin{center}
\resizebox{5.7in}{!}{\includegraphics{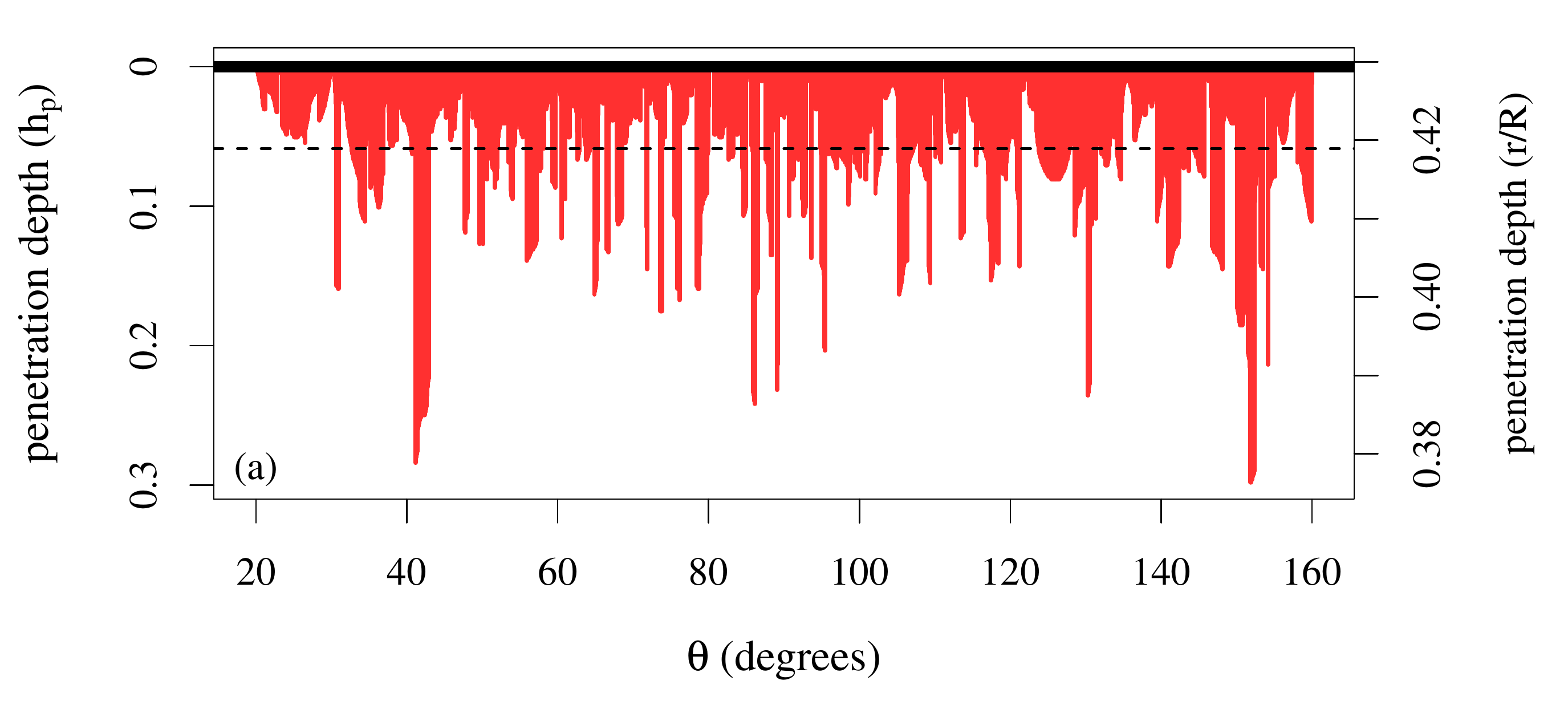}}
\resizebox{5.7in}{!}{\includegraphics{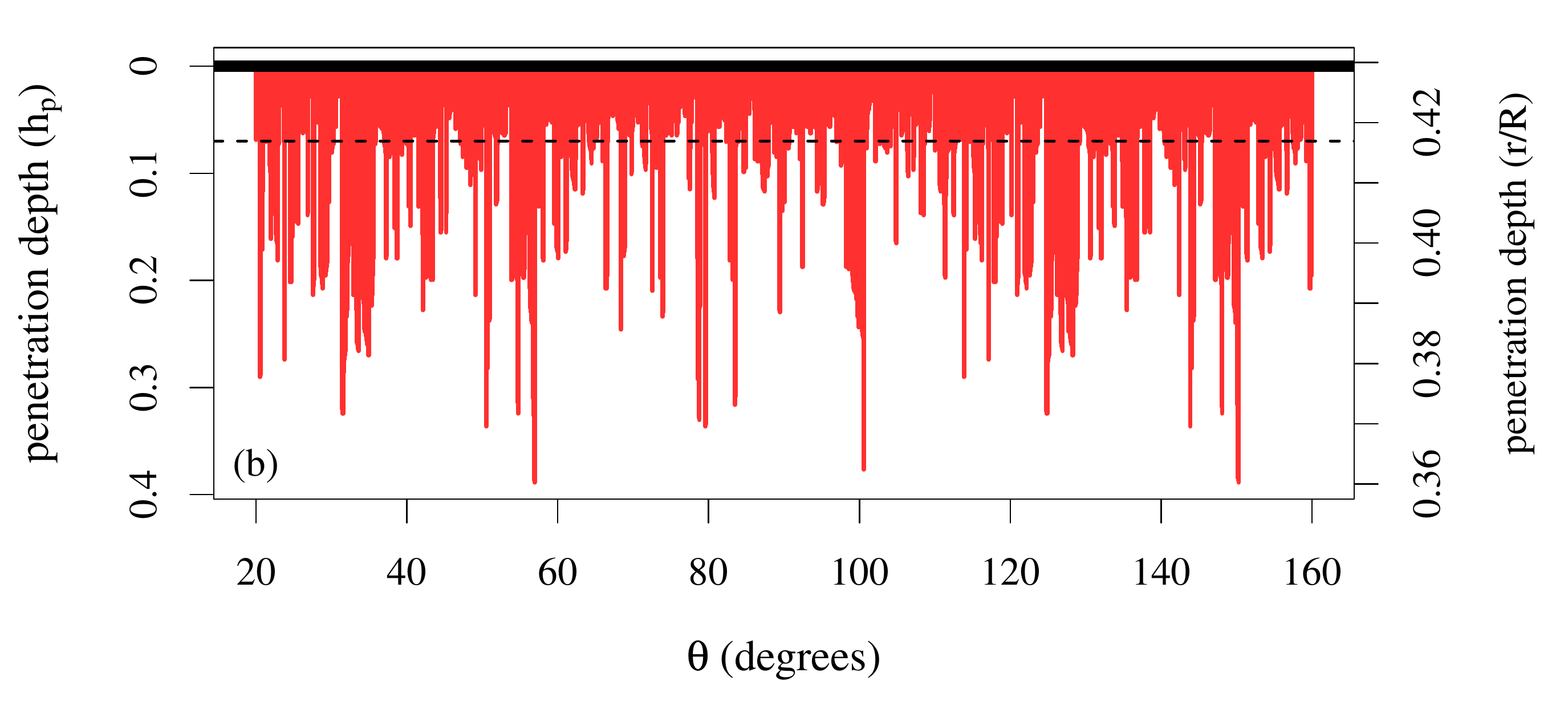}}
\resizebox{5.7in}{!}{\includegraphics{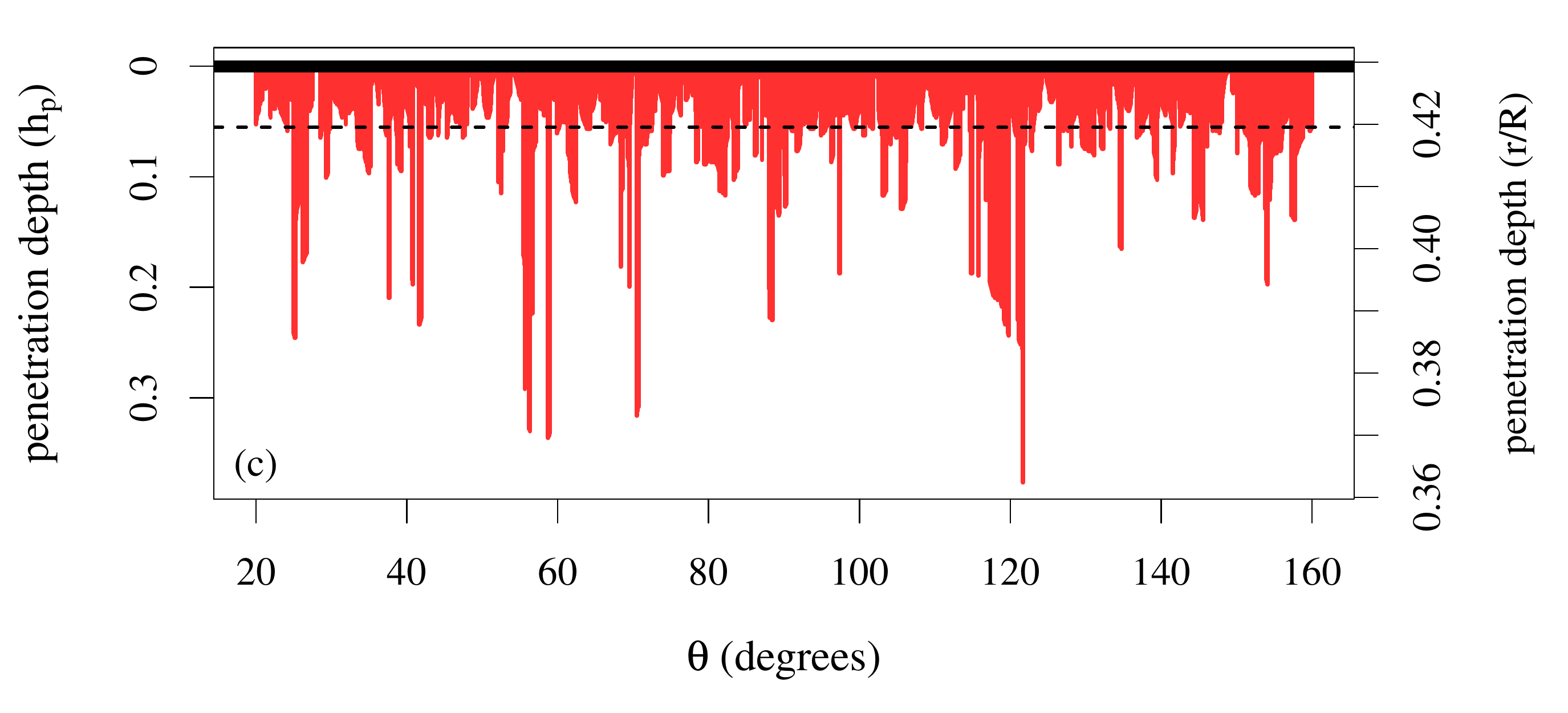}}
\caption{Angular structure of the penetration layer at three arbitrary times spread throughout simulation YS5.  The penetration depth in this illustration is determined by zeros of the vertical kinetic energy flux.  The boundary between the convection zone and the stable radiative zone is indicated by a solid black line.   The vertical axis is in units of the pressure scale height $h_p$ at this boundary.  A dashed black line indicates the average penetration depth at this time.
\label{figdiagram}}
\end{center}
\end{figure*}

Using the first zero of the vertical kinetic energy flux and the vertical heat flux in the stable zone as our criteria, we examine the depth of convective penetration for simulations YS0-5, described in Table~\ref{simsuma}.  In hydrodynamic simulations without rotation, the penetration is statistically independent of the colatitude $\theta$.  However at any instant in time the depth that plumes penetrate beyond the convection zone varies in $\theta$, defining the structure of the penetration layer.  Fig.~\ref{figdiagram} shows three typical snapshots that illustrates the shape of plumes penetrating into the radiative zone.  In the figure, the bottom of the convection zone, established by the entropy profile and the Schwarzschild criterion, is shown as a thick horizontal line.  The depth of convective penetration at any angular position in the spherical shell, based on the zero point of the vertical kinetic energy flux, is shown as a shaded area below this line.  At the angular resolution of simulation YS5, 
penetrating plumes often fill several grid spaces.
A general penetration depth calculated by a straightforward average in angle is indicated by the dashed horizontal line in each case; indeed we find that in Fig.~\ref{figdiagram}(a), for example, an average will hide the physically important more deeply penetrating plumes at approximately $40 \degree$ and approximately $150 \degree$.   This figure is calculated from the vertical kinetic energy flux; a qualitatively similar picture is produced from the vertical heat flux. We see similar pictures for each of the simulations considered in this work, independent of the resolution or boundary conditions used.  Comparing with three-dimensional hydrodynamic simulations during steady convection, we also observe similar pictures, although statistics based on long-times are not available for three-dimensional stellar convection.

\begin{figure*}
\begin{center}
\resizebox{3.5in}{!}{\includegraphics{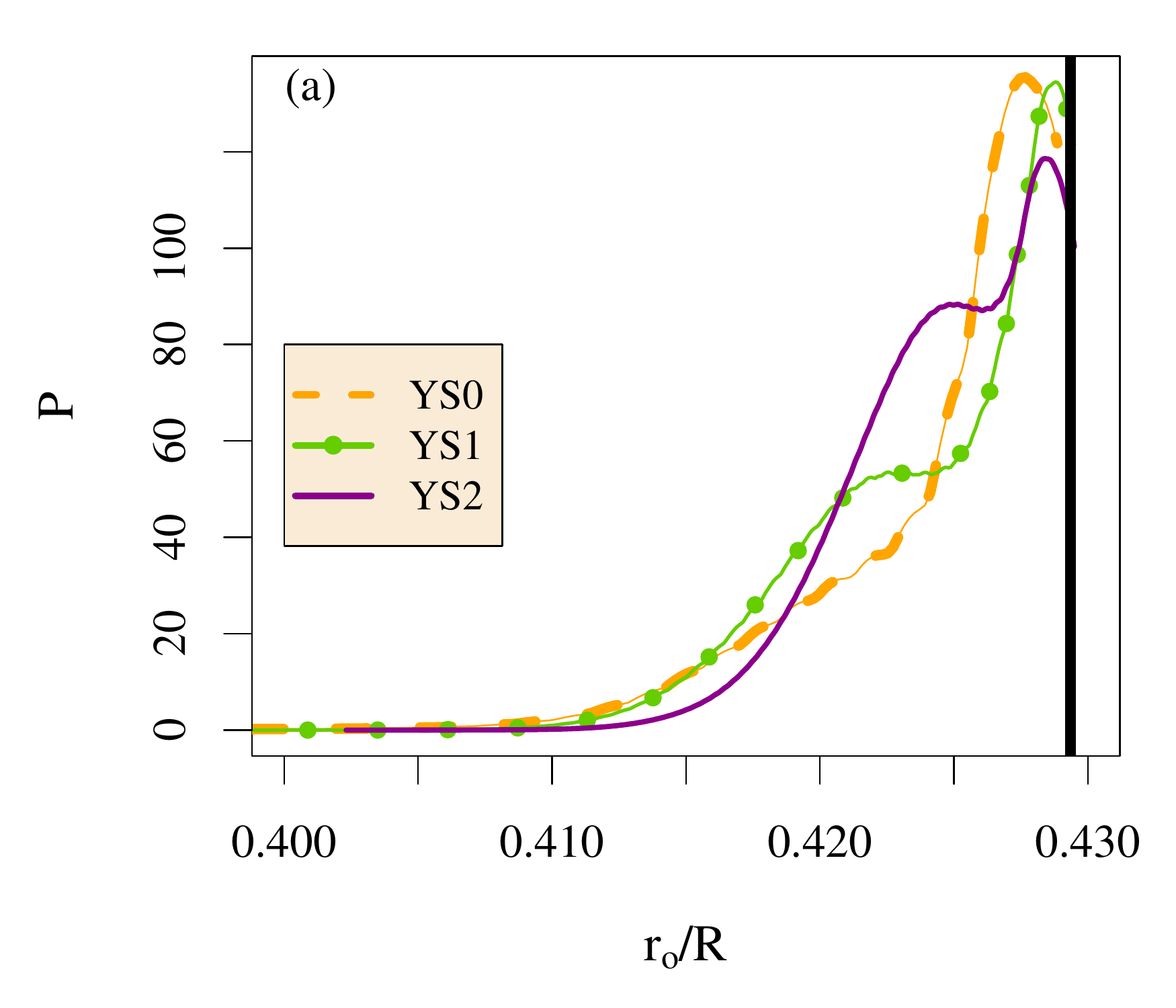}}\resizebox{3.5in}{!}{\includegraphics{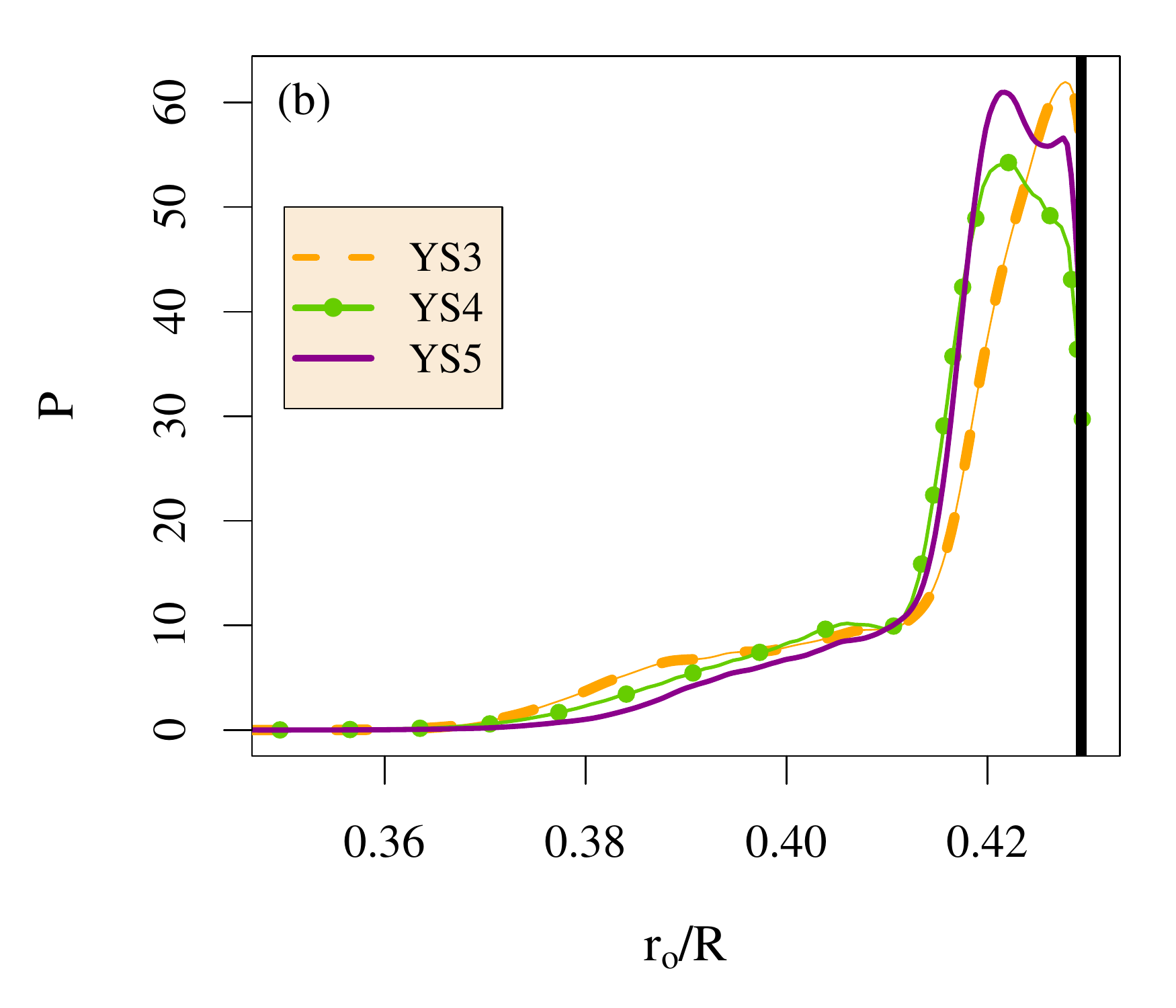}}
\resizebox{3.5in}{!}{\includegraphics{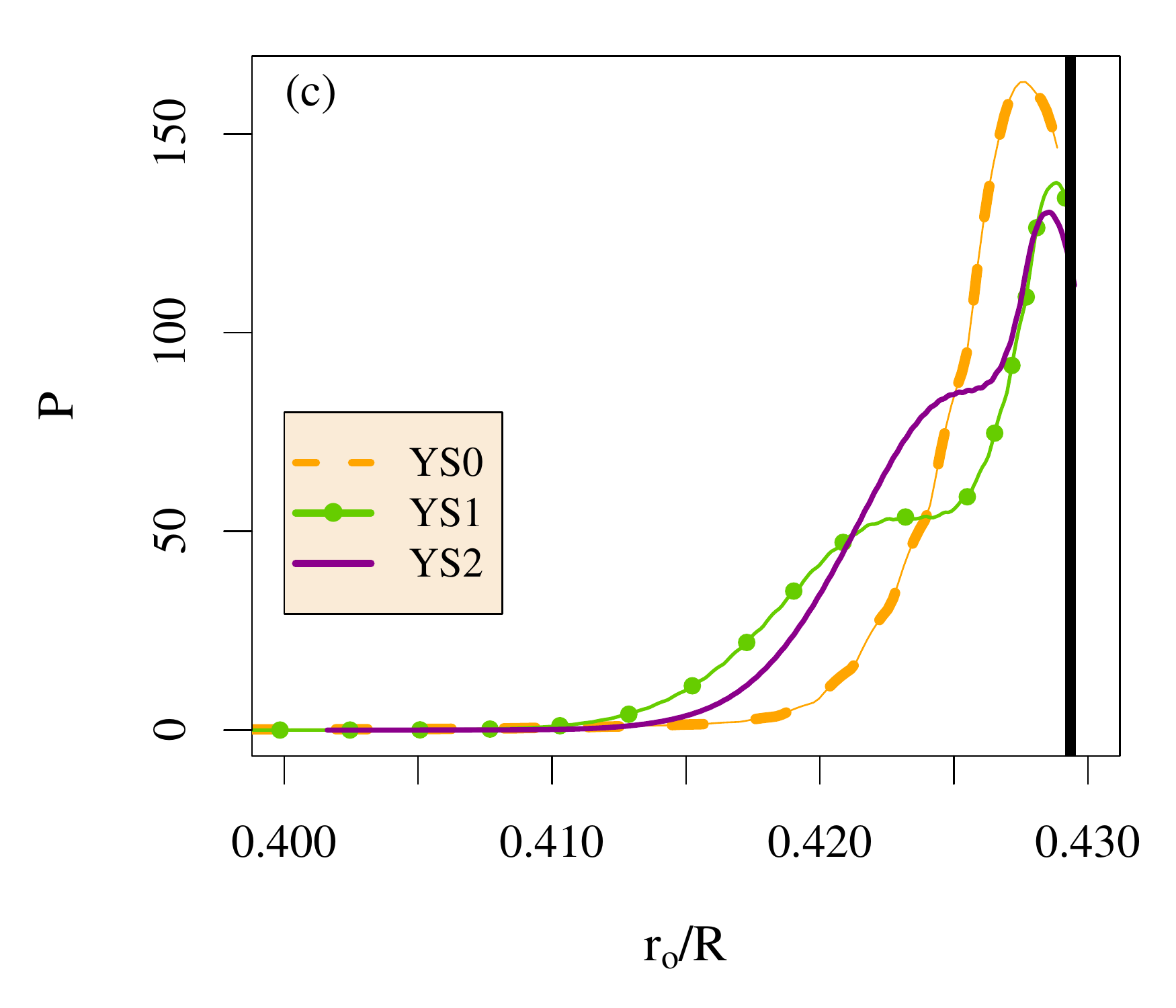}}\resizebox{3.5in}{!}{\includegraphics{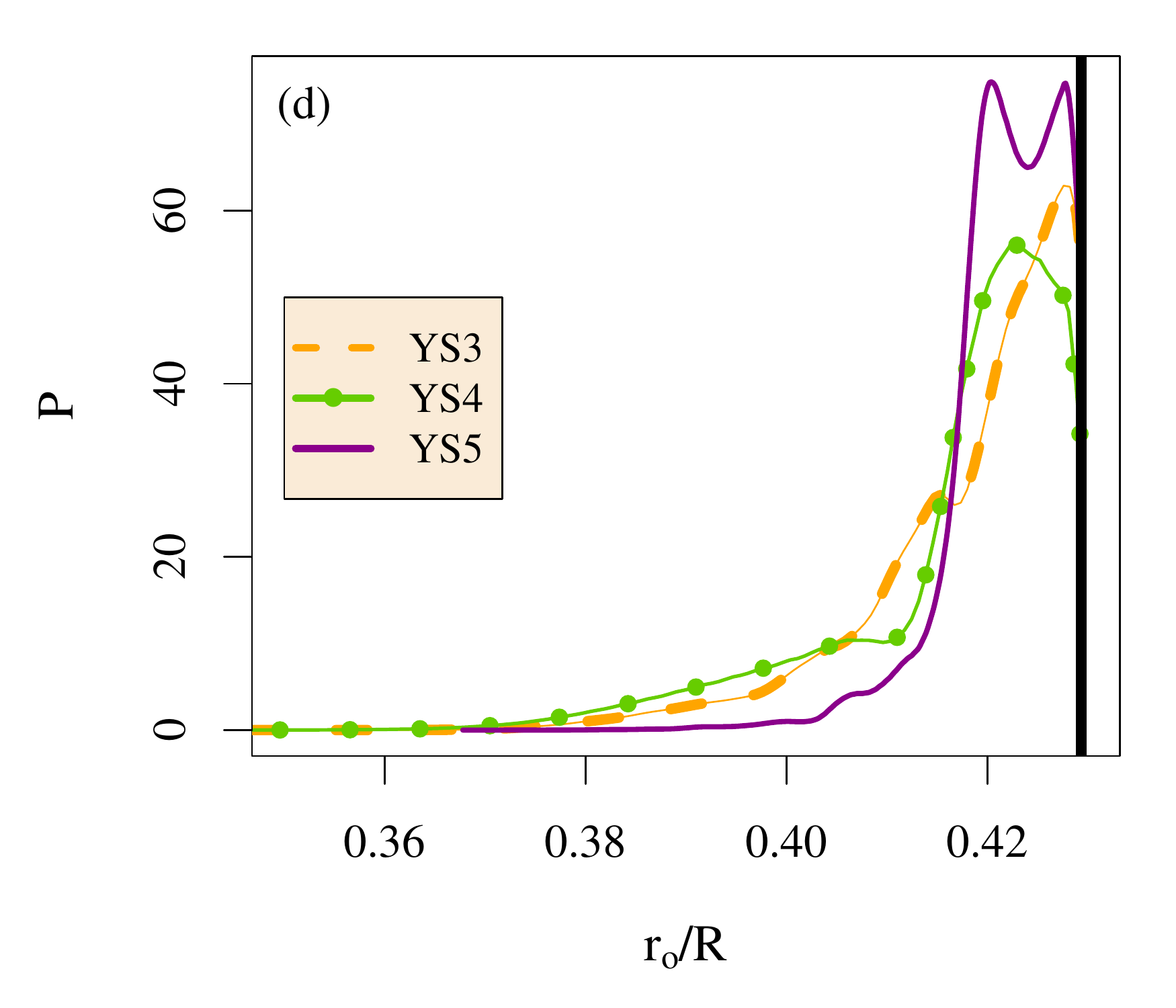}}
\caption{The PDF of the penetration position, $r_{\mathsf{o}}$, based on plumes penetrating at all angles and sampled at a fixed time interval throughout the full time span of each simulation indicated in Table~\ref{simsuma}.  (a) Penetration depth calculated from the vertical kinetic energy flux for the local convection simulations YS0-2.  (b) Penetration depth calculated from the vertical kinetic energy flux in the non-local convection simulations YS3-5.  (c) Penetration depth calculated from the vertical heat flux in the local convection simulations YS0-2.  (d) Penetration depth calculated from the vertical heat flux in the non-local convection simulations YS3-5.    The heavy vertical black line marks the convective boundary.
\label{figallovershootpdfs}}
\end{center}
\end{figure*}

To build a statistical model of convective penetration, we first consider the extent that plumes penetrate into the radiative zone at each angle defined in our simulation grid, and we sample the simulation data at a fixed time interval identical for each simulation.  This fixed time interval is on the order of $\tau_{\mathsf{conv}}/10^3$, a value selected to capture the fastest moving plumes entering the penetration layer.  We define the penetration position $r_{\mathsf{o}}$ as the position where convective motions cease based on either of our criteria. Fig.~\ref{figallovershootpdfs} shows the probability density function (PDF) of this penetration position for each set of simulations.  Each PDF is calculated using bins of equal size, which are identical for each set of simulations, and is properly normalized.    From these PDFs, two important observations can be made.
First, the PDFs possess either a multi-modal shape or a significant \emph{shoulder}. This identifies two distinct layers within the region below the bottom of the convection zone.  Immediately below the convection zone is a shallow layer where weak plumes frequently penetrate the radiative zone.  Below this layer is a second layer where stronger plumes penetrate, and more dramatically affect the stable radiative zone.  This strong penetration occurs with a typical probability between approximately $1/2$ and $1/7$th of the shallow layer.  The second important observation arises from the probability of larger penetration: these PDFs have a heavy tail that is non-Gaussian in appearance.  Both of these observations suggest that a straightforward average does not accurately describe the data.

Unlike the averaged profiles, the PDFs obtained based on the zeroes of the vertical kinetic energy flux and the zeroes of the vertical heat flux in  Fig.~\ref{figallovershootpdfs} yield characteristically similar results.
The PDFs calculated from the vertical heat flux are slightly more sharply peaked for small penetration extent than those calculated from the vertical kinetic energy flux.  However, the peaks and shoulders are present and in approximately the same places.

Meaningful differences between the PDFs calculated for the local simulations YS0-2 and the non-local simulations YS3-5 are clear from Fig.~\ref{figallovershootpdfs}. The PDF of the non-local simulations show that penetration in simulations YS3-5 reaches deeper into the radiative zone.  In addition, the multi-model behavior of the PDFs and the heavy tail is more significant.  This may be directly related to the higher velocities that arise for identically-resolved simulations when a large portion of the star is simulated, as reported by \citet{pratt2016spherical}.   Additional effects, such as the larger range of spatial scales of plumes that appear in the non-local simulations, may also be responsible, but are more difficult to quantify.

\section{Extreme value statistics of maximal convective penetration \label{secevd}}

\subsection{Formulation}
\begin{figure}[h]
\begin{center}
\resizebox{3.5in}{!}{\includegraphics{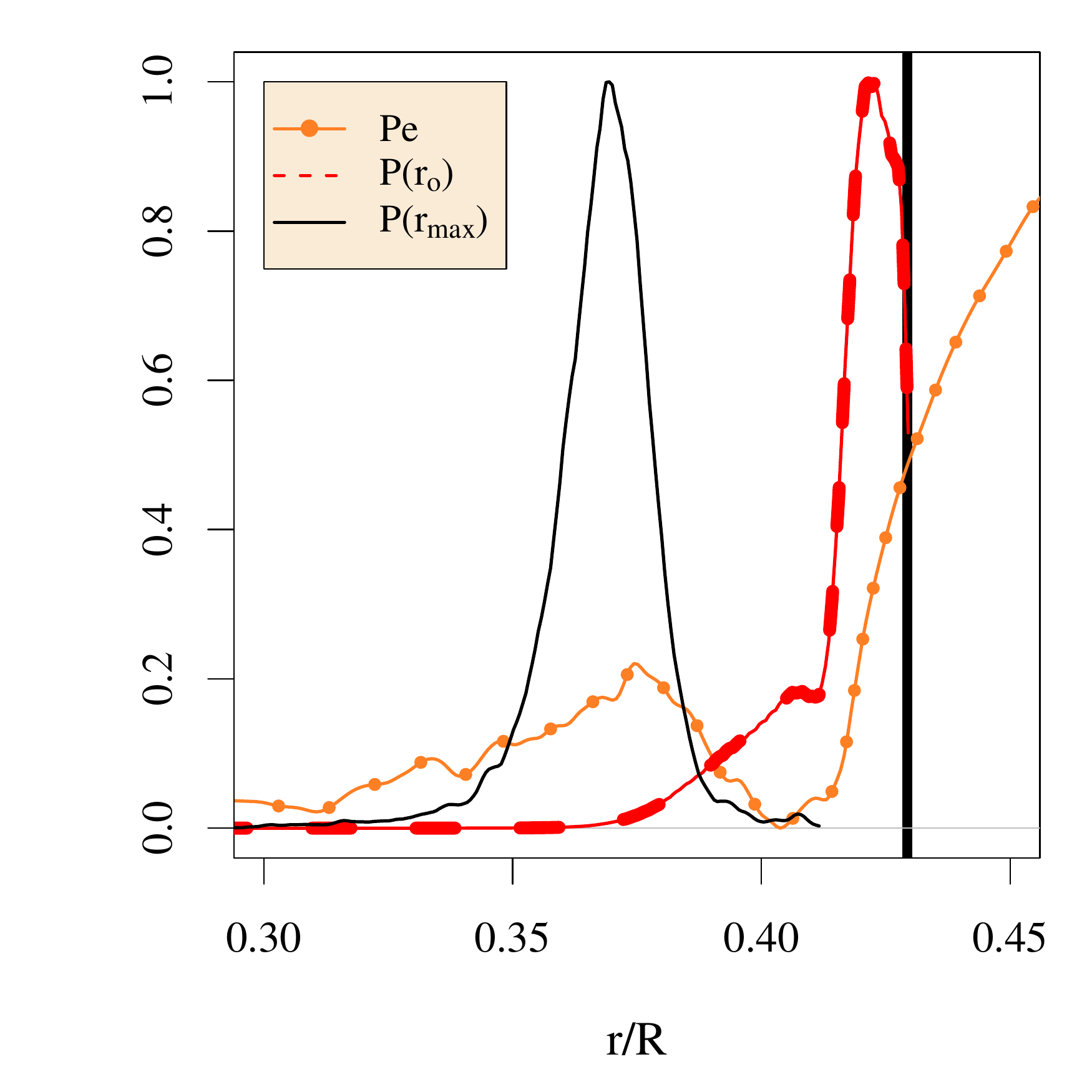}}
\caption{PDF of the convective penetration depth for all plumes $r_{\mathsf{o}}$ (dashed line), PDF of the maximal penetration depth at any time $r_{\mathsf{max}}$ (solid line), and the time- and horizontally volume-averaged P\'eclet number (dotted line) for simulation YS4. Each of these quantities is vertically shifted and scaled for comparison.  The convective boundary is delineated by a heavy vertical line.  
\label{figjustifyextremes}}
\end{center}
\end{figure}

The heavy tails evident in Fig.~\ref{figallovershootpdfs} motivate a more detailed statistical examination to characterize the depth of convective penetration that is meaningful for modeling stellar evolution.  To pursue this, we define a \emph{maximal penetration depth} to be the lowest position in the radiative zone that is reached at a given time:
\begin{eqnarray} \label{defrmax}
r_{\mathsf{max}} (t) = \mathsf{min}_{\theta} \left( r_{\mathsf{o}} (\theta, t) \right)  ~. 
\end{eqnarray}
Here we select the minimum over data at different angles, at the same time.
In Fig.~\ref{figjustifyextremes}  the PDF of this maximal penetration depth is compared with the PDF of the penetration depth of all plumes (as shown in Fig.~\ref{figallovershootpdfs}) and the time- and horizontally volume-averaged P\'eclet number.
The P\'eclet number reaches a minimum near the bottom of the convection zone, above $r/R=0.4$.  Between $0.35<r/R<0.4$, however, the P\'eclet number raises
again; this is the signature of waves excited by convective penetration, that form in a narrow band in radius below the convection zone in the young sun.  The RMS velocity profile and the RMS radial velocity profile are impacted by the velocity of the waves in this radial range.  The PDF of the maximal penetration depth correlates with the position where the P\'eclet number has a maximum due to these waves.    The quantity $r_{\mathsf{max}}$ thus pinpoints an important physical consequence of convective penetration, while the PDF of the penetration depth for all plumes $r_{\mathsf{o}}$ does not.

Because the PDF of the maximal penetration depth targets the waves that are the most obvious dynamical effect of penetration, it is reasonable to suggest that it should determine the relevant penetration length (or overshooting length) for stellar evolution calculations.  We therefore define a \emph{maximal penetration length} to be the difference between the maximal penetration depth and the convective boundary:
\begin{eqnarray} \label{defdeltar}
\Delta r_{\mathsf{max}} (t) = \mathsf{max}_{\theta} \left| r_{\mathsf{o}} (t) - r_{\mathsf{B}} \right| ~. 
\end{eqnarray}
Here $r_{\mathsf{B}}$ is the radial position of the convective boundary and the maximum is taken over all angles $\theta$ at a single time $t$.  Although in this work we treat penetration below a convection zone, the absolute value allows us to refer to penetration above or below a convective boundary.
This construction allows us to focus on the larger amount of convective penetration
that is likely to influence stellar structure.

The maximal penetration length $\Delta r_{\mathsf{max}}$ could be used straightforwardly as an overshooting length in simulations of stellar evolution.  
A simple average of eq.~\eqref{defdeltar} produces an overshooting length $\ell_{\mathsf{ov}} \sim 0.3 h_p$ in our simulations, where $h_p$ is the pressure scale height at the convective boundary.   A calculation of an overshooting length based on a time and horizontally averaged profile of the vertical kinetic energy flux, as in Fig.~\ref{kfluxplot}, produces a much smaller overshooting length, $\ell_{\mathsf{ov}} \sim 0.1 h_p$ from our simulations.  In recent years stellar calculations have adopted overshooting lengths in a wide range $0.05 h_p \leq \ell_{\mathsf{ov}} \leq 0.4 h_p$ \citep{basu1997seismology,chen2002low,brummell2002penetration,rogers2006numerical,politano2010population,liu2012three,montalban2013testing,jin2015convection}, which includes both of these values.  Both of these values are consistent with previously reported results for simulations with a similar stiffness to the young sun \citep{brummell2002penetration,
rogers2006numerical}.  Those earlier results demonstrate that the overshooting length is dependent on the Prandtl number, the Rayleigh number, and the resolution of the simulation, as well as the stiffness.   However our analysis suggests that larger values of the overshooting length than previously predicted would better reproduce the physics of large P\'eclet number convective penetration in the interior of a young low-mass star.

\subsection{Application of Extreme Value Theory}

The maximal penetration length in eq.~\eqref{defdeltar} suggests concepts from extreme value theory \citep{castillo2005extreme}.  In extreme value theory, unlike the central limit theorem that underlies much of probability theory, the interest is not the central values of an underlying distribution, but in accurately modeling the tails.  The centerpiece of extreme value theory is the generalized extreme value distribution (GEVD), 
which can be used to model the probability of maximal events.   The cumulative distribution function (CDF) identified with the GEVD has the form:
\begin{eqnarray}\label{eqgevd}
F(x) = \exp{ \left(-\left(1+\kappa \left( \frac{x-\mu}{\lambda} \right) \right)^{-1/\kappa} \right)} ~.
\end{eqnarray}
Here $\kappa$ is commonly called the shape parameter, $\mu$ is the location parameter, and $\lambda$ is the scale parameter \citep[see e.g. Section 9.1.1 eqs. (9.3) and (9.4) of][]{castillo2005extreme,charras2013extreme,gomes2015extreme}.  
Historically the well-documented Frech\'et distribution corresponds to a positive shape parameter $\kappa>0$, the Weibull distribution corresponds to $\kappa<0$, and the Gumbel distribution corresponds to $\kappa=0$.  The shape parameter is related directly to the heaviness of the tail of the underlying distribution.  For this reason $\kappa$ is also sometimes called the extreme value index (EVI).
  For the Gumbel case of $\kappa=0$ the GEVD may be expressed in the simpler form:
\begin{eqnarray}\label{eqgevdGUMBEL}
F(x) = \exp{ \left[- \exp{ \left(- \left( \frac{x-\mu}{\lambda} \right)  \right)}\right]} ~.
\end{eqnarray}
Thus the graph of the natural log of the negative natural log of $F$ will be a straight line if the distribution is of the Gumbel form.   
It will curve downward if the distribution is of the Weibull form, and upward if the distribution is of the Frech\'et form.  
When a strict upper limit exists for the variable considered, the Weibull case is predicted.
 Our maximal penetration length is defined so that $0<\Delta r_{\mathsf{max}} < 0.43 R$, meaning that convective plumes cannot penetrate beyond the center of the star. Thus a Weibull distribution is expected.

\begin{figure*}
\begin{center}
\resizebox{3.5in}{!}{\includegraphics{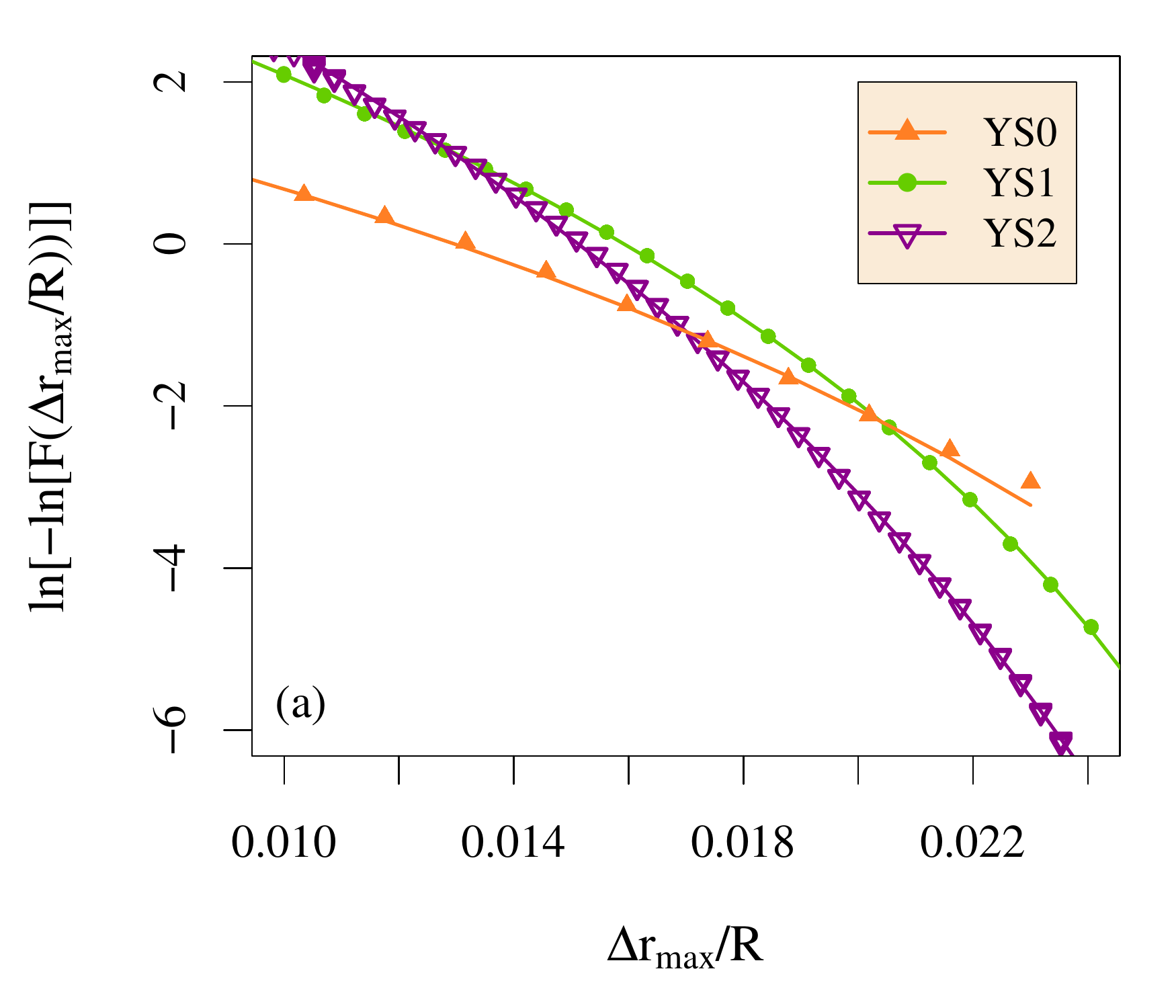}}\resizebox{3.5in}{!}{\includegraphics{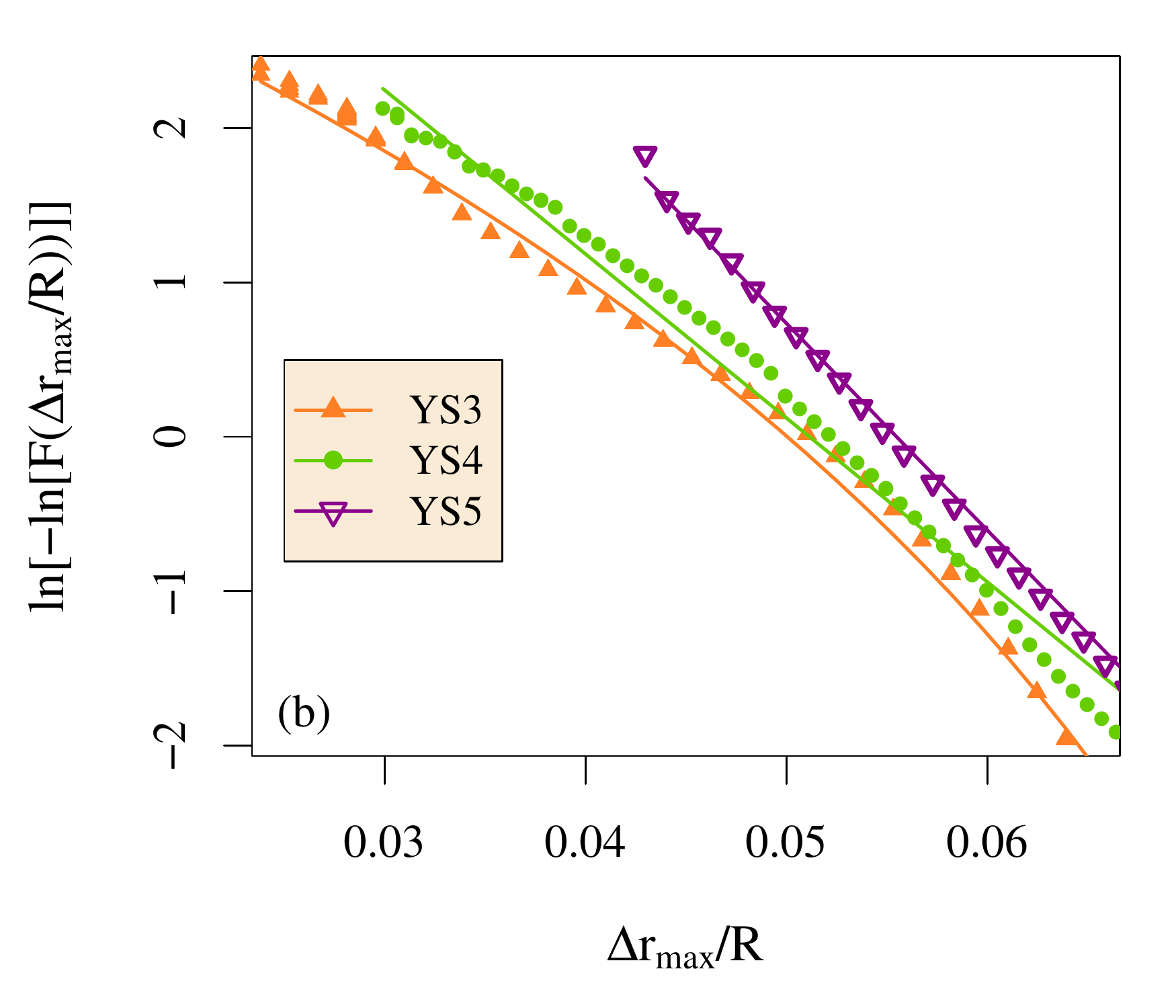}}
\resizebox{3.5in}{!}{\includegraphics{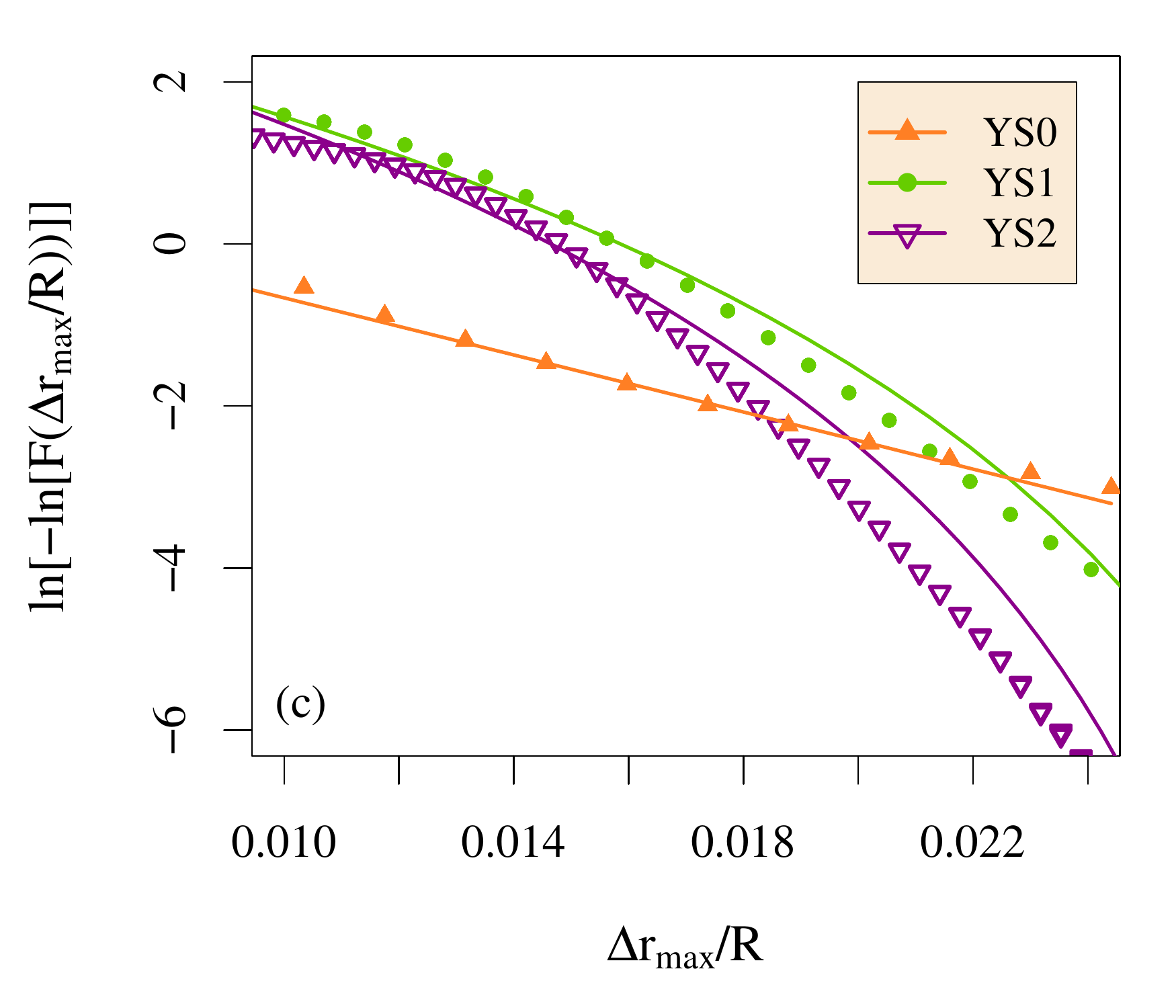}}\resizebox{3.5in}{!}{\includegraphics{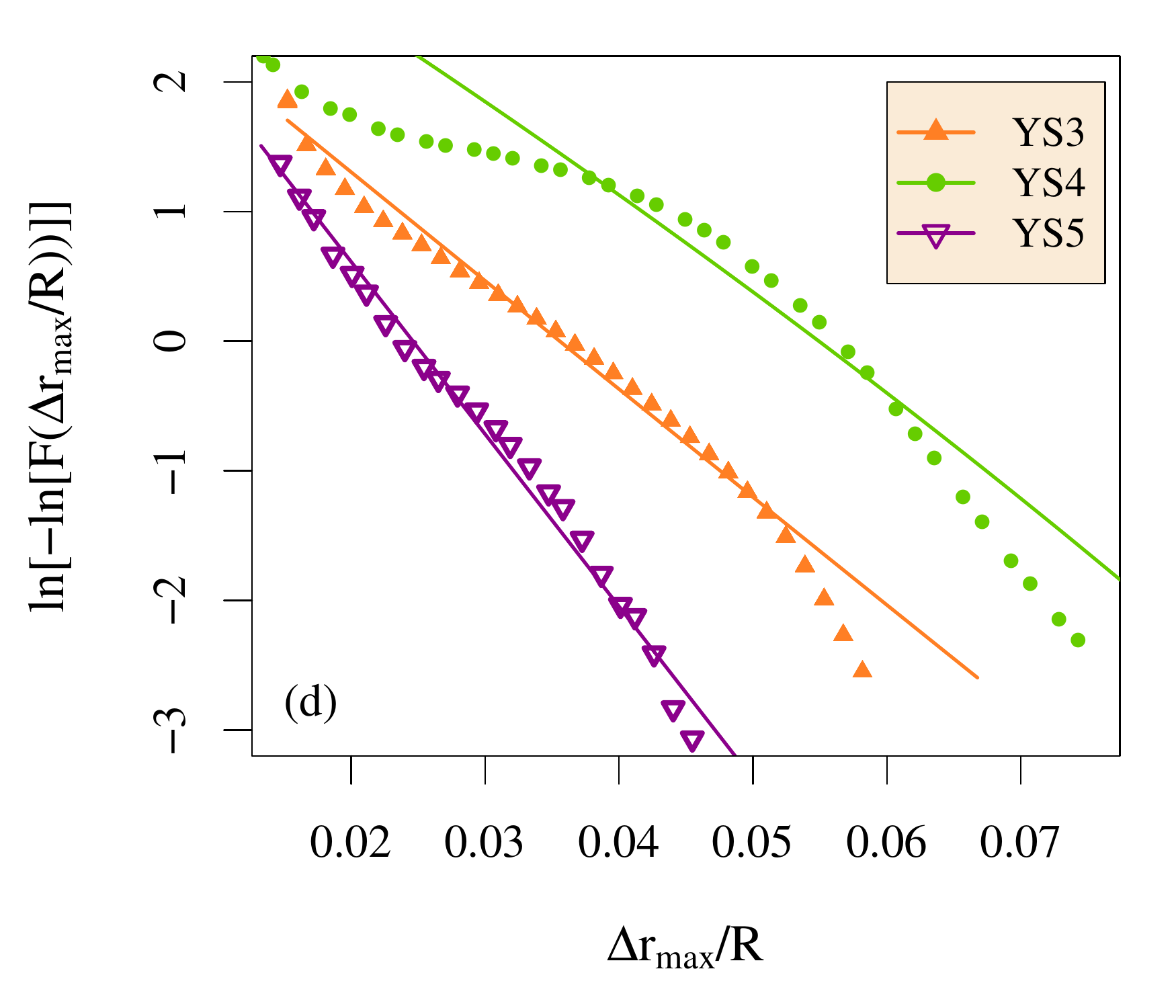}}
\caption{The cumulative distribution function $F$ of the maximal penetration length, $\Delta r_{\mathsf{max}}$ defined in eq.~\eqref{defdeltar}.  (a) Maximal penetration length calculated from the vertical kinetic energy flux for the local convection simulations YS0-2.  (b) Maximal penetration length calculated from the vertical kinetic energy flux for the non-local convection simulations YS3-5.  (c) Maximal penetration length calculated from the vertical heat flux for the local convection simulations YS0-2.  (d) Maximal penetration length calculated from the vertical heat flux for the non-local convection simulations YS3-5.  Data is shown as point symbols, while a fit to the GEVD form is shown as a line.
\label{figgumbeltest}}
\end{center}
\end{figure*}
Fig.~\ref{figgumbeltest} shows the CDFs of the maximum penetration length for our simulations.   
The forms of the CDFs in Fig.~\ref{figgumbeltest} indeed appear to have slight downward curvature, consistent with the Weibull form.
For all of these CDFs, most data lie at lower values of $\Delta r_{\mathsf{max}}$, and thus the distribution is clearest in this region; a lack of data at high values of $\Delta r_{\mathsf{max}}$, due to the comparative rarity of highly penetrating plumes, may contribute to the downward curvature in this region.   However simulations with different resolution produce remarkably similar CDFs, although different resolutions produce different convective velocities, and different spans of time are examined.   Increased resolution of a simulation leads to an improved characterization of the physical flows.  In a direct sense, it also leads to a larger sample size over which our maximum is taken to define the length $\Delta r_{\mathsf{max}}$.  However in a fluid simulation the data sampled is not strictly independent, so that it is not clear to what extent a higher resolution is related to an improvement in the use of extreme value theory, where taking a maximum over a large sample of independent data is assumed.  We do find that identical distributions are produced when our data is sampled ten times less frequently in time, on the order of the autocorrelation time of the flow at the bottom of the convection zone.

To precisely determine the parameters of the GEVD for each of our simulations, we use the package \emph{evd} \citep{revdpackage,penalva2013topics} publicly available for R (The R Project for Statistical Computing)\footnote{ The R Project for Statistical Computing:  https://cran.r-project.org/}.   Independently, we confirm these parameters using a nonlinear least squares fit to the GEVD.  Although in our fluid simulations the data is not strictly independent, as assumed by extreme value theory, we find that the GEVD provides an excellent fit.  
The shape, location, and scale parameters resulting from our fit are summarized in Table~\ref{tablekineticevd}.  In the table a value of $\kappa=0$ is indicated when the best fit is of the Gumbel form. The value of the shape parameter $\kappa$ is either small and negative, or zero for all of our simulations.    The available time-series of data for our highest resolution simulations is shorter than for lower resolutions. Despite this, a statistically significant trend is observed in which the value of $\kappa$ is explicitly zero in the high resolution, high velocity simulations YS4 and YS5.  Based on this observation, we predict that the Weibull characterization of our distributions should converge toward the simpler Gumbel distribution at high resolution, realistic stellar velocities, and when stellar convection is observed over  long times.

\begin{table*}
\begin{center}
\caption{Parameters for the generalized extreme value distribution of maximal convective penetration length $\Delta r_{\mathsf{max}}$.
 \label{tablekineticevd}
 }
\begin{tabular}{lccccccccccccccccccccccccccc}
\hline\hline
                      & location parameter $\mu$ & scale parameter $\lambda$ & shape parameter $\kappa$ %
\\ \hline \hline
vertical kinetic energy flux
\\ \hline \hline
YS0                 &  $1.34  \cdot 10^{-2} \pm  6 \cdot 10^{-6}$   &  $4.60 \cdot 10^{-3} \pm  ~3 \cdot 10^{-6}$ &   $ -0.133  \pm 6  \cdot 10^{-4}$ %
\\ \hline
YS1                  &  $1.63 \cdot 10^{-2} \pm  5 \cdot 10^{-6}$ &   $2.50 \cdot 10^{-3} \pm ~ 1 \cdot 10^{-6}$  &   $ -0.115  \pm 1 \cdot 10^{-3}$ %
\\ \hline
YS2                  &  $1.51 \cdot 10^{-2} \pm  1 \cdot 10^{-5}$ &  $1.84 \cdot 10^{-3} \pm ~ 3 \cdot 10^{-6}$ &    $ -0.10  \pm  3 \cdot 10^{-3}$%
\\ \hline \hline
YS3                  & $5.01  \cdot 10^{-2} \pm  4  \cdot 10^{-5}$  &  $ 8.87 \cdot 10^{-3} \pm ~3 \cdot 10^{-5}$  &  $-0.21  \pm 1 \cdot 10^{-3}$ %
\\ \hline 
YS4                  & $5.11  \cdot 10^{-2} \pm  7  \cdot 10^{-5}$  &  $9.40 \cdot 10^{-3} \pm ~4 \cdot 10^{-5}$  &  0  %
\\ \hline 
YS5                  &  $5.55  \cdot 10^{-2} \pm  1 \cdot 10^{-4}$   & $7.46 \cdot 10^{-3} \pm ~8 \cdot 10^{-5}$    & 0  %
\\ \hline \hline
YS6                  &  $6.34    \cdot 10^{-2} \pm  1   \cdot 10^{-4}$  &  $1.71   \cdot 10^{-2} \pm ~1 \cdot 10^{-4}$   &  0 %
\\ \hline \hline

vertical heat flux
\\ \hline \hline
YS0                 &   $6.22 \cdot 10^{-3} \pm ~  5 \cdot 10^{-6} $   &  $5.68 \cdot 10^{-3} \pm  4 \cdot 10^{-6} $  &  0  %
\\ \hline
YS1                  &  $1.59 \cdot 10^{-2} \pm  ~ 6 \cdot 10^{-6}$      &   $3.13 \cdot 10^{-3} \pm  1 \cdot10^{-6} $ &   $ -0.21 \pm  1 \cdot 10^{-6}$  %
\\ \hline
YS2                  &  $1.46  \cdot 10^{-2} \pm   ~ 2 \cdot 10^{-6}$ &   $2.72 \cdot 10^{-3} \pm  2 \cdot 10^{-6} $ &  $ -0.20 \pm  1 \cdot 10^{-3}$   %
\\ \hline \hline
YS3                  &  $3.55  \cdot 10^{-2} \pm ~  5 \cdot 10^{-5}$   &   $1.20 \cdot 10^{-2} \pm 3 \cdot 10^{-5}$   &  0 %
\\ \hline 
YS4                  &  $5.49  \cdot 10^{-2} \pm ~  8 \cdot 10^{-5}$   &   $1.28 \cdot 10^{-2} \pm 5 \cdot 10^{-5}$     & 0  %
\\ \hline 
YS5                  &   $2.46  \cdot 10^{-2} \pm ~  1 \cdot 10^{-4}$   &   $7.52 \cdot 10^{-3} \pm 8 \cdot 10^{-5}$    & 0 %
\\ \hline \hline
YS6                  &   $3.70 \cdot 10^{-2} \pm  ~  3 \cdot 10^{-5}$     & $8.52  \cdot 10^{-3} \pm ~2 \cdot 10^{-5}$  & 0 %
\\ \hline\hline
\end{tabular}
\tablefoot{Parameters $\mu$ and $\lambda$ are given in units of $R$, the stellar radius.  The shape parameter $\kappa$ is nondimensional.  Errors provided are the standard error of the fit.}
\end{center}
\end{table*}

\subsection{Radiation with the local surface temperature \label{secbb}}

\begin{figure}[h]
\begin{center}
\resizebox{3.5in}{!}{\includegraphics{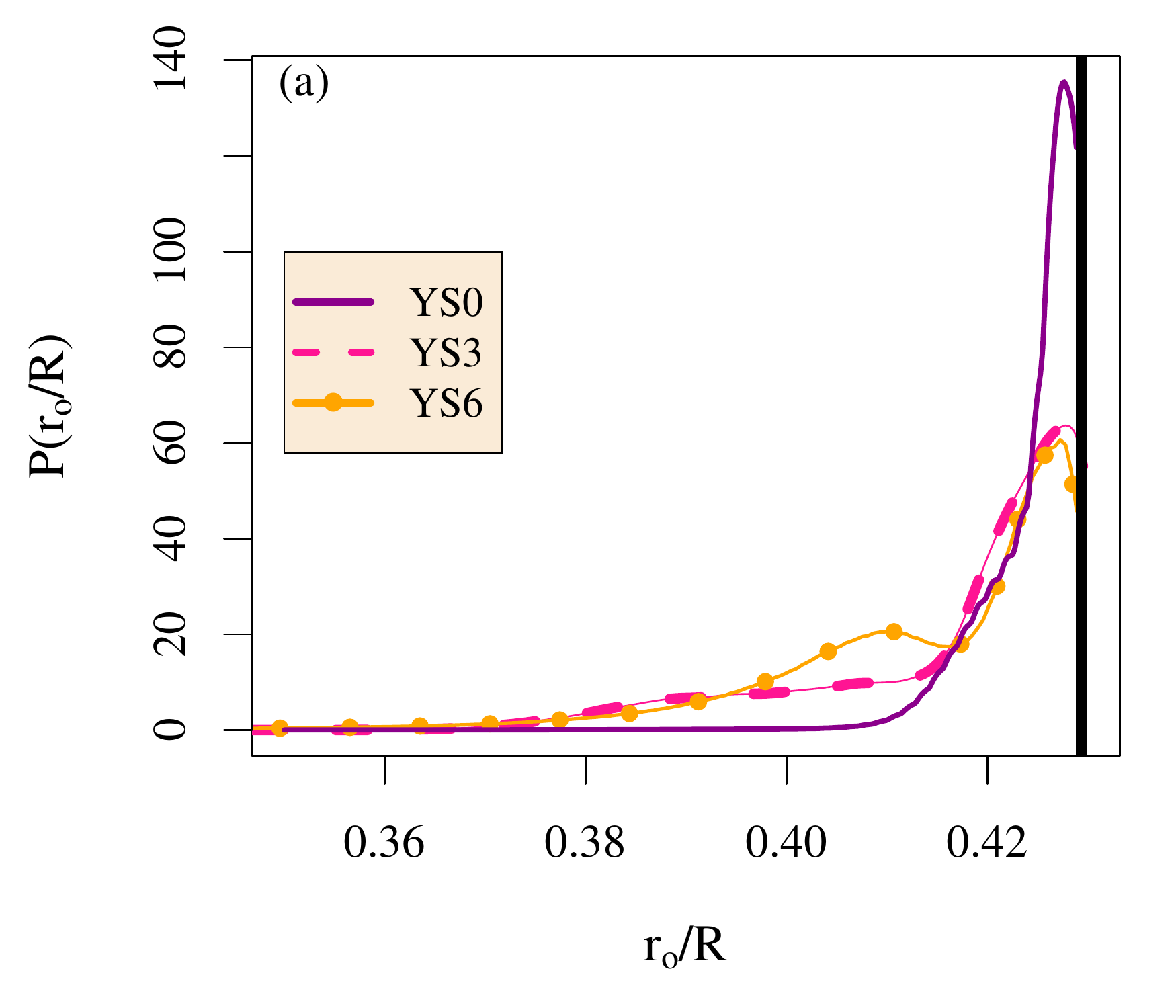}}
\resizebox{3.5in}{!}{\includegraphics{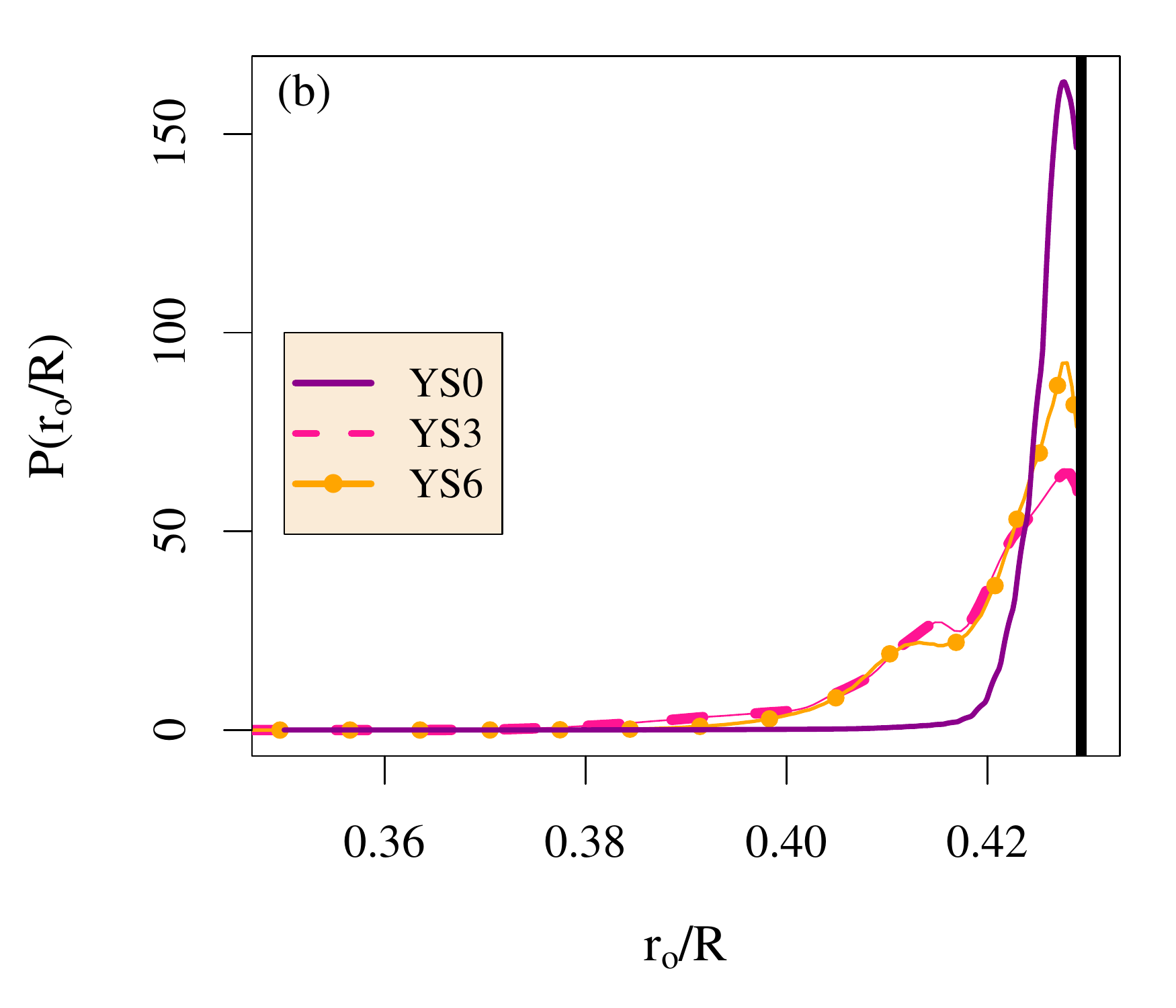}}
\caption{The PDF of the penetration extent of all plumes, $r_{\mathsf{o}}$.   (a) Penetration depth calculated from the vertical kinetic energy flux. (b) Penetration depth calculated from the vertical heat flux.  The convective boundary is delineated by a heavy vertical line.
\label{figbbovershootpdfs}}
\end{center}
\end{figure}

To understand whether the treatment of the surface produces different results for convective penetration, we perform an additional simulation YS6.  
This simulation has identical resolution to simulations YS0 (local convection) and YS3 (non-local convection), but in simulation YS6 the near-surface layers are included and the surface of the star is allowed to radiate energy with the local surface temperature.  
The PDF of the penetration depth of all plumes, $r_{\mathsf{o}}$, for YS6 is highly similar to simulation YS3; this comparison is shown in Fig.~\ref{figbbovershootpdfs}.  The penetration depth calculated from the vertical kinetic energy flux in Fig.~\ref{figbbovershootpdfs}(a) reveals a slightly larger probability of penetration between $0.39 < r < 0.42$ than in simulation YS3.  The penetration depth calculated from the vertical heat flux in Fig.~\ref{figbbovershootpdfs}(b) reveals a slightly larger probability of penetration between $0.42 < r < 0.43$ than in simulation YS3.  Although each of these simulations is observed over more than a hundred convective turnover times, the small difference in the statistics may still result from the limited length of time that these simulations are observed, particularly because convective penetration is intermittent.

The parameters for the GEVD of the maximal penetration length in simulation YS6 are included in Table~\ref{tablekineticevd}.  For both the vertical kinetic energy flux and the vertical heat flux criteria, the GEVD for simulation YS6 is best fit by a Gumbel distribution.  
The treatment of the surface in simulation YS6 is physically more accurate than in YS3.  That these two different treatments of the surface radiation produce similar results may be due to the fact that
both treat the surface only approximately and neglect important physical effects.
However the variable surface radiation in YS6 leads to higher velocity throughout the convection zone compared with either simulation YS3 or YS0; this was observed in \citet{pratt2016spherical}.
This is a significant observation, because it reinforces the finding that the overshooting length is not solely linked to the velocity of structures entering the penetration layer \citep{schmitt1984overshoot,hurlburt1994penetration,saikia2000examination,zahn2000plumes}.

\section{Determination of a diffusion coefficient for convection \label{secdiffcoef}}

\subsection{Background}

Convective overshooting and penetration have long been modeled by analyzing the enhanced diffusion in the penetration layer due to intermittent convective flows \citep[e.g. as described by][]{freytag2010role,noels2010overshooting,zhang2013convective}. Such models are based on the definition of a one-dimensional diffusion coefficient, as in
\begin{eqnarray}\label{diffeq}
\frac{\partial A(r,t)}{\partial t} =  \frac{1}{r^2}\frac{\partial}{\partial r} D r^2 \frac{\partial}{\partial r} A(r,t) ~.
\end{eqnarray}
Here $A$ is any scalar quantity, and $D$ is a diffusion coefficient.
Based on dimensional arguments, a diffusion coefficient has typically been estimated for astrophysical applications by either a characteristic length scale multiplied by a characteristic velocity scale \citep[e.g.][]{van1982overshoot,andrassy2015overshooting}, or a characteristic squared velocity multiplied by a characteristic time scale.

The diffusion equation~\eqref{diffeq} typically describes processes like molecular diffusion, and does not explicitly account for additional mixing due to large-scale convective motions, which enhanced diffusion.  In the setting of steady two-dimensional large-scale convection\footnote{No similar scaling is available in the more complex three-dimensional situation.}, the relation between the diffusion coefficient and the P\'eclet number of a convective flow has been studied in both the small and large P\'eclet number regimes \citep{moffatt1983transport,rosenbluth1987effective,shraiman1987diffusive,haynes2014dispersion}. In the small P\'eclet number regime this relation has been analytically and numerically shown to be:
\begin{eqnarray}\label{eqdiffpec1}
D(r) = D_0 \left(1 + c_1 \mathsf{Pe}(r)^2 \right) ~.
\end{eqnarray}
Here $D$ is the enhanced diffusion coefficient and $D_0$ is the diffusion coefficient based on small-scale diffusive processes, which may be molecular diffusion in a liquid, or small-scale turbulent diffusion in a fluid.  The constant $c_1$ is related to the aspect ratio of the convection rolls. 
In the large P\'eclet number regime the corresponding relation is:
\begin{eqnarray}\label{eqdiffpec2}
D(r) = c_2 D_0 \mathsf{Pe}(r)^{1/2} ~,
\end{eqnarray}
where $c_2$ is a constant based on the aspect ratio of the convection cells.  \citet{rosenbluth1987effective} evaluate $c_2$ for a wide range of possible convection roll aspect ratios, and find that $c_2 \sim \mathcal{O}(1)$ with only small variation.  Eqs.~\eqref{eqdiffpec1} and \eqref{eqdiffpec2} are derived for an ideal setting rather than stellar convection, but are suggestive of diffusive scalings relevant to different P\'eclet number regimes.

Near the surface of a star, the small P\'eclet number limit is most appropriate; the P\'eclet number in simulations of this layer has been reported to be of order one.  In this setting, eq.~\eqref{eqdiffpec1} indicates that an enhancement of the diffusion coefficient  dependent on the squared P\'eclet number is reasonable.   If the thermal diffusivity and characteristic length scale are assumed to be approximately constant, the P\'eclet number is proportional to the velocity.  This situation can be related directly to the enhanced diffusivity proposed by \citet{freytag1996hydrodynamical}.      In their radiation hydrodynamics simulations, \citet{freytag1996hydrodynamical} observe exponential decay of the RMS radial velocity beyond the boundary of the convection zone.  They use this result to define a piecewise-continuous diffusion coefficient that is constant in the convection zone.  Beyond the convection boundary, the piecewise-continuous diffusion coefficient is defined using the Ansatz that it should decay exponentially as the velocity, i.e.
\begin{eqnarray}\label{freitageq}
D_{\mathsf{F96}}(r) = v_{\mathsf{B}}^2 \tau \exp{(-2(r_{\mathsf{B}}-r)/h_{v})} ~.
\end{eqnarray}
Here $v_{\mathsf{B}}$ and $r_{\mathsf{B}}$ are the RMS radial velocity at the convective boundary, and the position of the convective boundary, respectively.  The pressure scale-height $h_{p}$ at the convective boundary is used to define the characteristic time scale $\tau = h_{p}/v_{\mathsf{B}}$ and the velocity scale-height at the convective boundary $h_{v} = f h_{p}$, where $f$ is a constant of proportionality.
The exponentially decaying diffusion coefficient of eq.~\eqref{freitageq} has been incorporated in several stellar evolution codes, including MESA \citep{paxton2013modules}, ATON 3.1 \citep{ventura2008aton}, and GARSTEC \citep{weiss2008garstec}.

\subsection{Diffusion coefficient in the large P\'eclet number regime}

\begin{figure}[hb]
\begin{center}
\resizebox{3.5in}{!}{\includegraphics{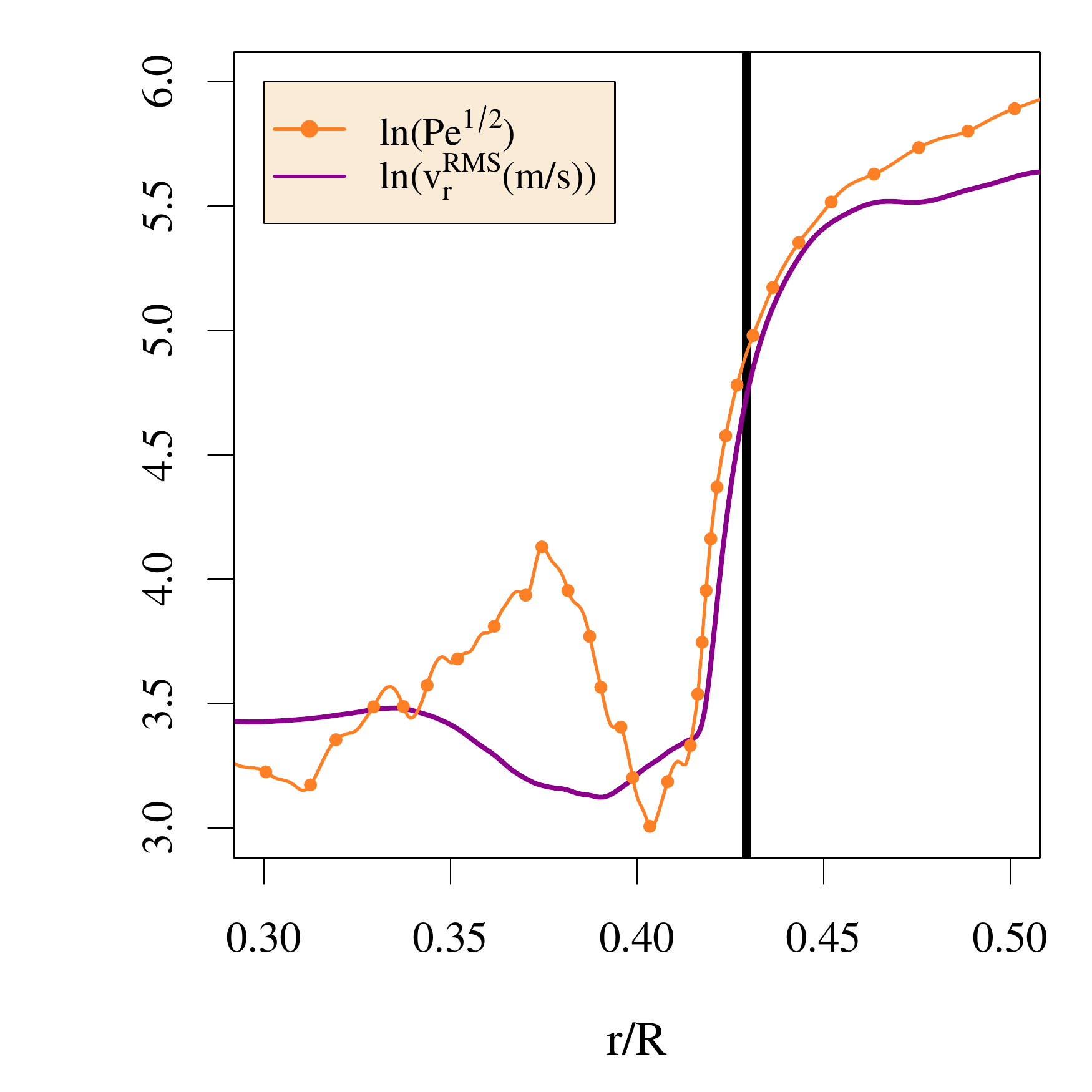}}
\caption{Time-averaged and horizontally volume-averaged RMS radial velocity near the bottom of the convection zone in simulation YS4.  The time-averaged and horizontally volume-averaged square-root of the P\'eclet number is shown, scaled and shifted, for comparison. The heavy vertical black line marks the convective boundary.
\label{fignotexponential}}
\end{center}
\end{figure}
In contrast to the early simulations examined by \citet{freytag1996hydrodynamical}, in our comparatively high-resolution young sun simulations the lower convective boundary is in the large P\'eclet number regime. For this physically and numerically distinct case, we observe that the RMS radial velocity does not decay exponentially beyond the convective boundary.  The decay of the natural log of RMS radial velocity is shown in Fig.~\ref{fignotexponential}, and it does not decay linearly with radius in the penetration layer.   
Motivated by eq.~\eqref{eqdiffpec2}, we examine the square-root of the P\'eclet number in this figure in addition to the RMS radial velocity, and find that predictably it 
exhibits a trend of decay outside the convection zone similar to the RMS radial velocity.    The observed departure from exponential decay motivates our proposal of a new form for the diffusion coefficient for stellar interiors in the large P\'eclet number regime.

\begin{figure}[h]
\begin{center}
\resizebox{3.5in}{!}{\includegraphics{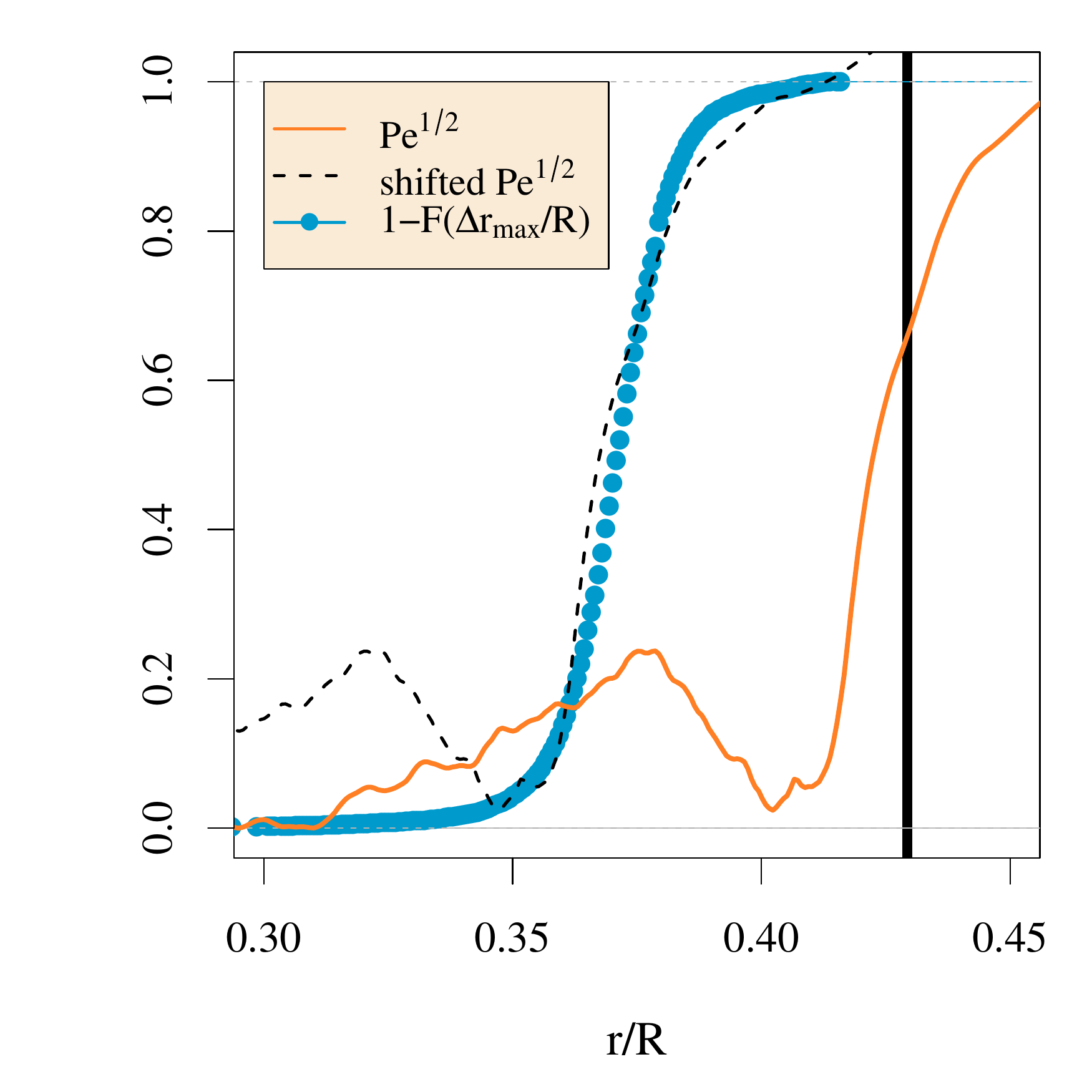}}
\caption{Time and horizontally volume-averaged square-root of the P\'eclet number compared with the data for the cumulative distribution function $F(r_{\mathsf{max}}/R) = 1-F(\Delta r_{\mathsf{max}}/R)$ calculated from the vertical kinetic energy flux.  The best fit to this data  from simulation YS4 is a Gumbel distribution with the parameters indicated in Table~\ref{tablekineticevd}.  The P\'eclet number has been normalized and shifted vertically.  The dashed line indicates a P\'eclet number that is also shifted in radius so that the common shape is clear.  The heavy vertical black line marks the convective boundary. 
\label{figdiffalternative}}
\end{center}
\end{figure}
\begin{figure}[h]
\begin{center}
\resizebox{3.5in}{!}{\includegraphics{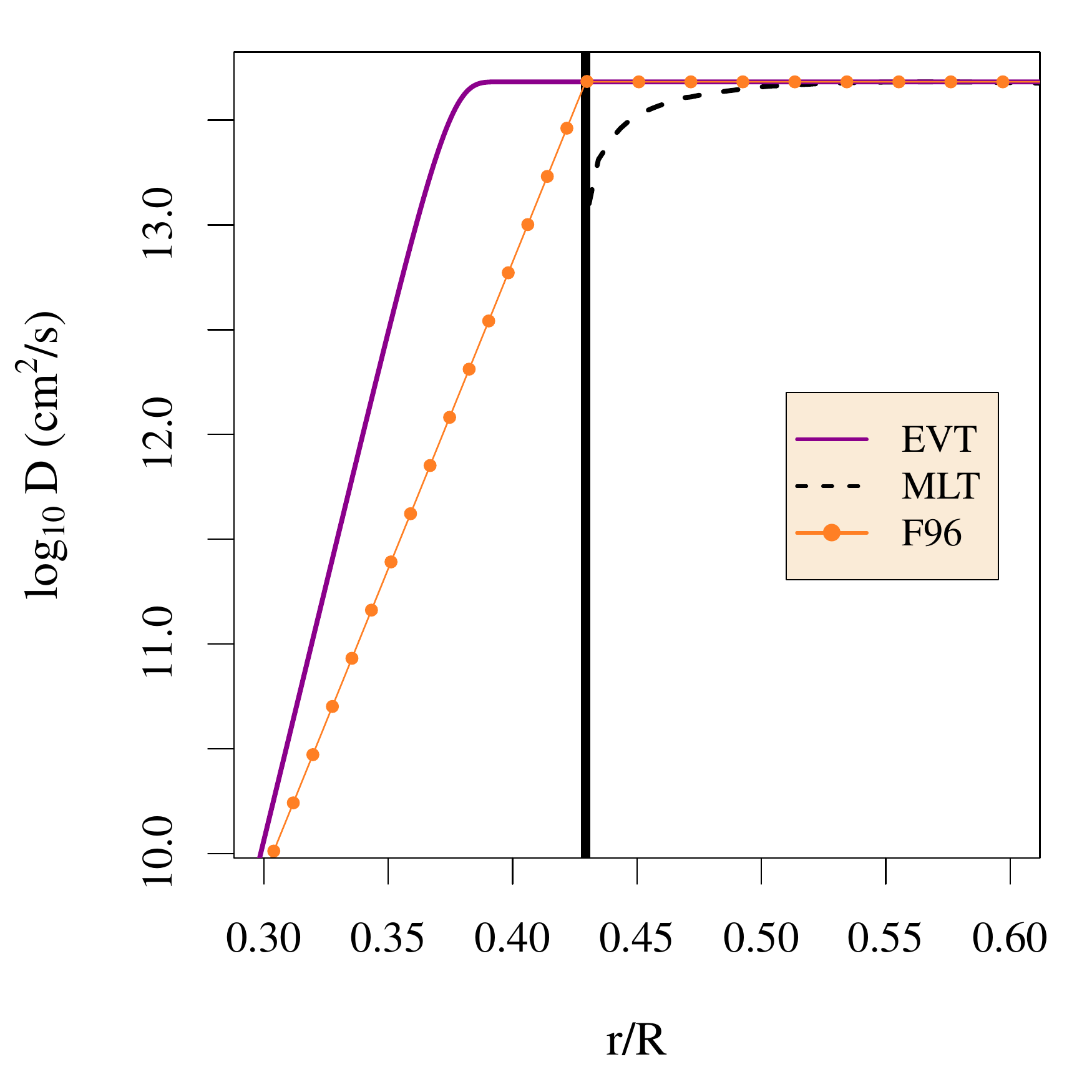}}
\caption{The diffusion coefficient based on EVT proposed in eq.~\eqref{newEVTdiffusion} compared with a diffusion coefficient defined from mixing length theory $D_{\mathsf{MLT}}$, and the decaying exponential diffusion coefficient $D_{\mathsf{F96}}$ of \citet{freytag1996hydrodynamical} defined in eq.~\eqref{freitageq}.  Here the peak of the mixing length theory diffusion coefficient has been used to calibrate diffusion in the convection zone.
\label{figdiffcomp}}
\end{center}
\end{figure}

Based on our statistical analysis in Section~\ref{secevd}, we predict that the maximal penetration length should match the profile of the Gumbel distribution: $ \exp{(-\exp{(-x)})}$.
 Fig.~\ref{figdiffalternative} shows that the radial profile of the square-root of the P\'eclet number closely matches the form of the CDF of maximal penetration length.  
 We propose a diffusion coefficient of the form:
\begin{eqnarray}\label{newEVTdiffusion}
D(r) = D_0 \mathsf{Pe_{B}}^{1/2} \left( 1- \exp{\left( - \exp{\left(-   \frac{(r_{\mathsf{B}}-r) - \mu}{ \lambda} \right)} \right)}  \right)~.
\end{eqnarray}   
Here penetration beneath a convection zone is assumed, and $\mathsf{Pe_{B}}$ is a characteristic P\'eclet number in the convection zone.  In constructing this diffusion coefficient, we have used the identity:
 \begin{eqnarray}\label{fid}
F(r_{\mathsf{max}}/R) = 1-F(\Delta r_{\mathsf{max}}/R)~,
\end{eqnarray}
 to relate the cumulative distribution functions of the maximal penetration depth $r_{\mathsf{max}}$ and the maximal penetration length $\Delta r_{\mathsf{max}}$.  This identity allows us to express the diffusion coefficient in the same form as \citet{freytag1996hydrodynamical}, as a difference between the radial coordinate and the position of the convective boundary indicated by $r_{\mathsf{B}}$. 
The prefactor $D_0 \mathsf{Pe_{B}}^{1/2}$ in eq.~\eqref{newEVTdiffusion} is representative of diffusion in the convection zone, and may be estimated by comparison with stellar evolution models for convective diffusion.  Using mixing length theory, diffusion in the convection zone is characterized by $D_{\mathsf{MLT}} = 1/3~L_{\mathsf{MLT}}~v_{\mathsf{MLT}}$, where the length scale is proportional to the pressure scale height\footnote{The 1D model for the young sun is produced using a mixing length $L_{\mathsf{ MLT}}=1.5 ~h_p$, which produces a diffusion coefficient of $D_{\mathsf{ MLT}} \approx 5 \cdot 10^{13} \mathsf{cm^2/s}$ in the convection zone.}.   The values for $\mu$ and $\lambda$ in eq.~\eqref{newEVTdiffusion} may be estimated from simulations YS3-5 in Table~\ref{tablekineticevd}.  In units of the pressure scale height  at the base of the convection zone, $h_{p} \approx 0.178 ~R $, these parameters give a range of
\begin{eqnarray}\label{usefulapprox}
\mu &\approx& 0.035~\mbox{--}~ 0.05 R = 0.2 ~\mbox{--}~ 0.3 ~h_{p}~,
\\
\lambda &\approx& 0.007 ~\mbox{--}~ 0.01 R = 0.04 ~\mbox{--}~ 0.055~ h_{p}~.
\end{eqnarray}
We note that $\mu$ is approximately equivalent to the maximal penetration length estimated in Section~\ref{secevd}.  
As with $D_{\mathsf{F96}}$, the diffusion coefficient in eq.~\eqref{newEVTdiffusion} should be applied only in the layer of the star affected by convective penetration, so that it does not introduce infinitesimally small amounts of mixing into deeper layers of the star.
This proposed diffusion coefficient is shown in Fig.~\ref{figdiffcomp}, calibrated using the mixing-length-theory diffusion coefficient, $D_{\mathsf{MLT}}$, used to produce the young sun model.  Both  $D_{\mathsf{MLT}}$, and $D_{\mathsf{F96}}$ are shown for comparison.  The mixing-length-theory diffusion coefficient does not include any diffusion in the penetration layer.  
A detailed comparison of the exponentially decaying diffusion coefficient $D_{\mathsf{F96}}$ with our diffusion coefficient in eq.~\eqref{newEVTdiffusion} naturally depends on the parameters that define the exponential decay.  These are commonly defined by treating the constant of proportionality between the velocity scale-height and pressure scale-height as 
 a free parameter $f \leq 0.1$ \citep[e.g.][]{angelou2015diagnostics}.
For these small values of $f$ our diffusion coefficient models a higher level of diffusion throughout the penetration layer than the exponentially decaying diffusion
coefficient of $D_{\mathsf{F96}}$.  Because of the higher level of diffusion predicted, and the different form of its decay below the convective boundary, 
it is worth exploring the effect of this new diffusion coefficient on the abundance of light elements like lithium and beryllium in solar-type stars \citep{boesgaard1976stellar,pinsonneault1994stellar}, and to characterize its observational signatures. In simulations that include magnetic fields, this differing amount of diffusion due to convective penetration may affect magnetic flux in the penetration layer, and thus have consequences for the stellar dynamo \citep[e.g.][]{nordlund1992dynamo,rudiger1995solar,tobias2001transport,rempel2003thermal,brummell2010dynamo}.

\section{Summary and Discussion \label{secfinal}}

We have studied flows that penetrate below a convection zone due to large-scale two-dimensional stellar convection in a prototypical young low-mass star. 
We have analyzed and compared convective penetration in two fundamental types of simulations: (1) those that model local convection by truncating the spherical shell to a minimal radial layer around the convective boundary, and (2) those that model non-local convection by simulating nearly the whole star.
  For each of these situations we examine stellar convection at three different resolutions, allowing us to observe penetration relating to a range of characteristic convective velocities. 
To compare with our simulations that hold the energy flux constant at the outer boundary, we also consider a non-local convection simulation where the surface radiation is allowed to vary with the surface temperature.  This simulation is characterized by higher velocities in the convection zone.  Our simulations do not account for rotation, shear flows, chemical mixing, or magnetic fields.  These processes may change the properties of the flow at the convective boundary and alter the quantitative results. Analysis of these effects will be explored in  future work.

Historically, a time and horizontally volume-averaged radial profile of the vertical kinetic energy flux has often been used to determine the depth of the penetration layer.
Different results are produced from a comparison of the time and horizontally volume-averaged radial profiles of the vertical kinetic energy flux with the vertical heat flux, a second way to define the penetration layer.  Here we show that when plumes at all angles and sampled at a fixed time interval are considered, PDFs of convective penetration depth calculated from these two different fluxes are characteristically similar.  The PDFs are non-Gaussian, with a heavy tail.  Based on these PDFs, we define two distinct layers that form between the convection zone and the stable radiative zone: a shallow layer where convective plumes frequently penetrate, and a deeper layer where convective plumes penetrate intermittently.   

Motivated by the non-Gaussian character of convective penetration, we examine the maximal penetration depth at any time.
 The PDF of this quantity is centered closely around the position where waves are excited by convective penetration in the radiative zone. This suggests that the maximal penetration depth is a physically significant length scale to define the penetration layer.  Based on a statistical analysis of our simulation data, we determine that a reasonable estimate of the overshooting length in the young sun is $\ell_{\mathsf{ov}} \sim 0.3 h_p$.  We observe that the cumulative distribution function of the maximal penetration length can be modeled accurately by a Weibull distribution with a small shape parameter.  As both higher resolution and higher velocities are examined, this Weibull distribution appears to converge toward a Gumbel distribution.   This approach toward measuring convective penetration is new.  Further work is required to demonstrate that it accurately describe convective penetration when rotation, shear flows or magnetic fields are considered.  However, it provides a promising avenue to quantify the mixing due to overshooting or penetration in stars.  Analysis of the properties of the waves generated by convective penetration into the radiative zone will be explored in future work.

Building on these statistical results, as well as scalings for two-dimensional convection, we propose a new form for the convection-enhanced diffusion coefficient suitable for one-dimensional stellar evolution calculations.   Unlike the $D_{\mathsf{F96}}$ diffusion coefficient \citep{freytag1996hydrodynamical}, our proposed diffusion coefficient is targeted for the large P\'eclet number flows characteristic of stellar interiors.  
This represents a step forward for stellar evolution modeling, because the $D_{\mathsf{F96}}$ diffusion coefficient
has been broadly applied to both small and large P\'eclet number flows, outside of its range of demonstrated physical validity.  Follow-up studies to explore the effect of our diffusion coefficient in one-dimensional stellar evolution calculations are now required to analyze possible differences that may result (Baraffe et al. in prep).  Isolating observable signatures to discern the structure of overshooting and penetration layers is currently a challenge, but may also shed light on the differences between these two forms of the diffusion coefficient.

In this work we have focused on identifying the amount of mixing due to convective penetration, which can be modeled by a simple diffusion coefficient.  However convective penetration also has a significant impact on the thermal profile of a penetration layer.  A complete model of penetration therefore should include  a treatment of heat transport relevant for stellar evolution calculations.  Such a model is currently being pursued based on the simulations presented.

\FloatBarrier
\begin{acknowledgements}
The research leading to these results has received funding from the European Research
Council under the European Union's Seventh Framework (FP7/2007-2013)/ERC grant
agreement no. 320478.
\\
I.B thanks MOCA at Monash University for warm hospitality.
\\
This work used the DiRAC Complexity system, operated by the University of Leicester IT Services, which forms part of the STFC DiRAC HPC Facility (www.dirac.ac.uk). This equipment is funded by BIS National E-Infrastructure capital grant ST/K000373/1 and STFC DiRAC Operations grant ST/K0003259/1. DiRAC is part of the National E-Infrastructure.
\\
This work also used the University of Exeter Supercomputers, ZEN and ISCA, funded by the STFC/DiRAC, the Large Facilities Capital Fund of BIS and the University of Exeter.
\end{acknowledgements}

\bibliographystyle{aa}
\bibpunct{(}{)}{;}{a}{}{,}
\bibliography{music}

\end{document}